\documentclass[journal]{IEEEtran}
\usepackage{amsmath,amsfonts}
\usepackage{algorithm}
\usepackage{algorithmicx}
\usepackage{algpseudocode}
\usepackage{textcomp}
\usepackage{stfloats}
\usepackage{url}
\usepackage{verbatim}
\usepackage{graphicx}
\usepackage{cite}
\let\originalappendices\appendices
\renewcommand{\appendices}{\originalappendices}
\usepackage{amsmath}

\usepackage{amsthm}

\usepackage{pgfplots}
\usepackage{graphicx}       
\usepackage{float}
\usepackage{subcaption}
\usepackage{mathrsfs}
\usepackage{multirow}
\usepackage{xcolor}  
\usepackage{booktabs}       
\usepackage{nicefrac}       
\usepackage{microtype}      
\usepackage{tabularx}
\usepackage{array}     
\usepackage{stfloats}
\usepackage{fontawesome}
\usepackage{amssymb}
\usepackage{pifont}
\usepackage{enumitem}

\newtheorem{lemma}{Lemma}

\pdfoutput=1

\floatname{algorithm}{Algorithm}

\begin{document}

\title{SFIBA: Spatial-based Full-target Invisible Backdoor Attacks}

\author{Yangxu Yin, Honglong Chen,~\IEEEmembership{Senior Member,~IEEE,}
Yudong Gao, Peng Sun, \\Zhishuai Li, Weifeng Liu,~\IEEEmembership{Senior Member,~IEEE}

\thanks{Yangxu Yin, Honglong Chen, Yudong Gao, Zhishuai Li and Weifeng Liu are with the College of Control Science and Engineering, China University of Petroleum (East China), Qingdao 266580, P. R. China. E-mail: yinyangxv@163.com; chenhl@upc.edu.cn; YudongGao0504@163.com; zhishuai.li@upc.edu.cn; liuwf@upc.edu.cn.}
\thanks{Peng Sun is with the College of
Computer Science and Electronic Engineering, Hunan University, Changsha 410082,
P. R. China. E-mail: psun@hnu.edu.cn.}
\thanks{Corresponding author: Honglong Chen.}
}



\maketitle


\begin{abstract}
Multi-target backdoor attacks pose significant security threats to deep neural networks, as they can preset multiple target classes through a single backdoor injection.
This allows attackers to control the model to misclassify poisoned samples with triggers into any desired target class during inference, exhibiting superior attack performance compared with conventional backdoor attacks.
However, existing multi-target backdoor attacks fail to guarantee trigger specificity and stealthiness in black-box settings, resulting in two main issues. First, they are unable to simultaneously target all classes when only training data can be manipulated, limiting their effectiveness in realistic attack scenarios. Second, the triggers often lack visual imperceptibility, making poisoned samples easy to detect.
To address these problems, we propose a \textbf{S}patial-based \textbf{F}ull-target \textbf{I}nvisible \textbf{B}ackdoor \textbf{A}ttack, called SFIBA. 
It restricts triggers for different classes to specific local spatial regions and morphologies in the pixel space to ensure specificity, while employing a frequency-domain-based trigger injection method to guarantee stealthiness.
Specifically, for injection of each trigger, we first apply fast fourier transform to obtain the amplitude spectrum of clean samples in local spatial regions. Then, we employ discrete wavelet transform to extract the features from the amplitude spectrum and use singular value decomposition to integrate the trigger.
Subsequently, we selectively filter parts of the trigger in pixel space to implement trigger morphology constraints and adjust injection coefficients based on visual effects, achieving dynamic, invisible, and effective trigger injection.
We conduct experiments on multiple datasets and models. 
The results demonstrate that SFIBA can achieve excellent attack performance and stealthiness, while preserving the model’s performance on benign samples, and can also bypass existing backdoor defenses.

\end{abstract}

\begin{IEEEkeywords}
Deep learning, multi-target backdoor attack, attack stealthiness, black-box settings.
\end{IEEEkeywords}

\begin{figure}[!t]
    \centering
    \includegraphics[width=0.46\textwidth]{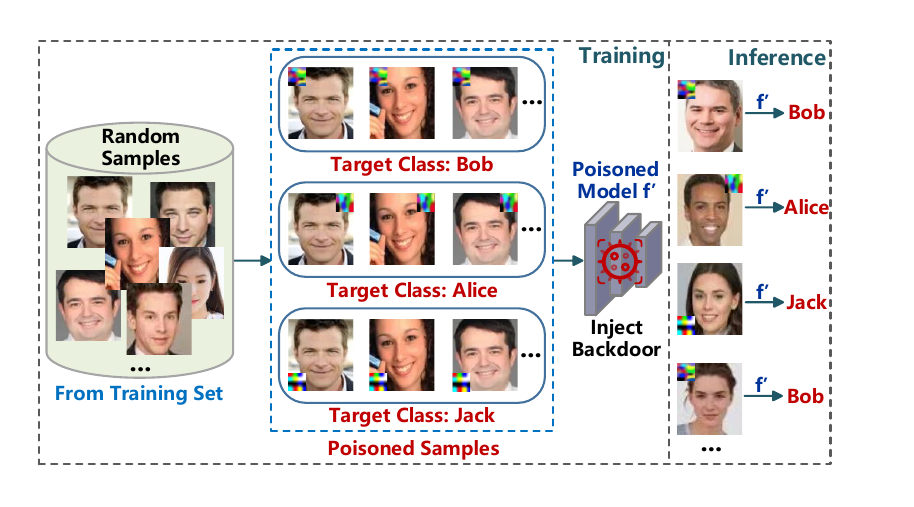}
    \vspace{-1mm}
    \caption{Schematic of multi-target backdoor attack. }
    \label{fig0:env}
    \vspace{-8mm}
\end{figure}

\section{Introduction}
\IEEEPARstart{I}{n} recent years, Deep Neural Networks (DNNs) has become integral to our daily lives due to their exceptional performance.
However, DNNs suffer from a lack of transparency and interpretability, which makes them susceptible to malicious attacks \cite{r35,r36}, including adversarial examples \cite{r1,r2}, poisoning attacks \cite{r3,r4}, and backdoor attacks \cite{r10,r5,r6,r7,r8}, etc. 
Among these, backdoor attacks require injecting backdoor into the model during training by modifying the training set \cite{r10,r5,r6,r7} or altering model parameters \cite{r8,r9}. Consequently, the model performs normally on clean samples, but misclassifies those with triggers into target class during inference. 
While backdoor attacks on image classification tasks have grown increasingly sophisticated, the majority of these attacks are single-target, meaning they can only designate one specific class as the target. 
In contrast, multi-target backdoor attacks \cite{r11,r12,r13,r14} feature broader payloads, allowing the presetting of multiple target classes. 
Here, ``payload" refers to the number of classes that can be simultaneously attacked, and each target class is mapped to a specific trigger injection method. 
This enables attackers to flexibly control the classification of poisoned samples into any desired predefined target class during inference, rather than a single fixed target class, as shown in Fig. \ref{fig0:env}.


Moreover, research on multi-target backdoor attacks is essential.
In certain scenarios, attackers may need to switch the target class after injecting a backdoor, which is common in real-world situations. 
For example, attackers might aim to bypass facial recognition models to access company backends. With conventional single-target backdoor attacks, they can only make the model classify poisoned images as a specific person, like employee Bob. However, if Bob leaves and his data is removed, the backdoor no longer works, forcing attackers to retrain the model to embed a new backdoor, which is a tedious and time-consuming task. In contrast, multi-target backdoor attacks with powerful payloads allow attackers to flexibly switch target classes after a single backdoor injection. This means that if Bob leaves, attackers can simply switch the target to another employee, Alice, without retraining, ensuring continued access.
Therefore, multi-target backdoor attacks with powerful payloads pose a significant threat to deep models, attracting much attention from both academia and industry.


However, existing multi-target backdoor attacks face a significant limitation: they struggle to attack all classes (i.e. full-target backdoor attacks) while ensuring trigger stealthiness in black-box settings. To elaborate, attacking all classes means establishing mappings between each class and its respective class-specific trigger injection method, which is necessary to achieve powerful backdoor payloads.  
In black-box settings, attackers can only manipulate the training set and have no knowledge of the victim model's architecture or parameters, nor can they interfere with the model training process. 
Furthermore, the black-box settings above are highly representative of real-world attack scenarios.
Thus, the ability to maintain powerful backdoor payloads under such constraints is essential for ensuring the successful execution of multi-target and even full-target backdoor attacks.
However, achieving this capability is difficult, as the primary challenge lies in ensuring the specificity of triggers for different classes without involving the victim model.
If trigger specificity is insufficient, injecting multiple backdoors targeting different classes can cause interactions that degrade backdoor performance and severely limit backdoor payloads.
In addition, ensuring trigger stealthiness is also important.
If trigger stealthiness is insufficient, the poisoned samples will be easily detected, leading to the failure of the attack.
So far, no attack paradigm has successfully ensured trigger specificity and stealthiness within the constraints of black-box settings to execute invisible attacks capable of targeting a full range of classes.

To address the above issues, we propose a \textbf{S}patial-based \textbf{F}ull-target \textbf{I}nvisible \textbf{B}ackdoor \textbf{A}ttack, called SFIBA, where we leverage the sensitivity of backdoor to trigger spatial locations and morphologies to ensure trigger specificity and provide a theoretical proof for the spatial sensitivity. Meanwhile, we design a frequency-domain-based trigger injection method to ensure both trigger stealthiness and effectiveness. 
Specifically, we divide samples into multiple disjoint blocks, referred to as $Block$s, and inject invisible triggers with specific morphological constraints into different $Block$s for various target classes.
When injecting the trigger, we first perform Fast Fourier Transform (FFT) on both the clean image within the $Block$ and the trigger image. This converts the trigger injection from the pixel space to the frequency domain, providing an initial guarantee of stealthiness.
Next, to address the challenges in selecting the trigger injection region and coefficient caused by $Block$ size limitations, we introduce Discrete Wavelet Transform (DWT) and Singular Value Decomposition (SVD).
In particular, we apply DWT to extract diagonal features from the clean amplitude spectrum and then inject the trigger's corresponding features by fusing singular values. This is followed by an inverse transformation back to the pixel space.
Then, we apply DWT again to the poisoned $Block$ and filter parts of the trigger to implement morphological constraints.
Finally, we dynamically adjust the injection coefficient of the trigger in each poisoned sample to ensure an excellent visual appearance. 
The spatial and morphological constraints of the trigger ensure that even similar trigger strengths do not significantly undermine its specificity.
Our contributions can be summarized as follows:
\begin{itemize}
    \item We propose a \textbf{S}patial-based \textbf{F}ull-target \textbf{I}nvisible \textbf{B}ackdoor \textbf{A}ttack (SFIBA), which can attack all classes in black-box settings, i.e., constructing class-specific trigger injection methods and establishing mappings between these methods and their corresponding target classes.
    \item We leverage the spatial and morphological constraints to ensure trigger specificity, and theoretically demonstrate the backdoor's spatial sensitivity. Furthermore, we design a frequency-domain-based method to balance stealthiness and effectiveness of the trigger located in a finite $Block$.
    \item We validate the effectiveness of SFIBA on various datasets and its robustness against advanced defenses, demonstrating that it can achieve high attack success rates for any predefined target class while maintaining model performance on benign samples.
\end{itemize}

The remainder of the paper is organized as follows. Section \ref{sec:related_work} reviews the related work on backdoor attacks and defenses. Section \ref{sec:Threat_Model} details the capability of attackers and attack modeling. Section \ref{sec:Methodology} delves into the methodology of SFIBA, including its motivation and key steps. Section \ref{sec:Evaluation} evaluates SFIBA's effectiveness, stealthiness, robustness, and performs ablation experiments. Finally, Section \ref{sec:Conclusion} concludes the paper.

\section{Related Work}
\label{sec:related_work}
\subsection{Backdoor Attacks}
\textbf{Single-Target Backdoor Attacks.}
Research on single-target backdoor attacks generally focuses on improving trigger stealthiness and effectiveness.
Early patch-based triggers \cite{r6} are visually detectable. Subsequent methods, such as blending \cite{r15}, reflection \cite{r16}, and warping \cite{r5}, enhance the stealthiness of triggers to a certain degree. Later triggers, designed based on frequency-domain methods like FFT (FIBA) \cite{r17} and discrete cosine transform (Ftrojan) \cite{r18}, achieve invisibility in the pixel space. In some recent studies \cite{r7,r19}, sample-specific triggers have been generated, making the triggers even more invisible and challenging to detect. Additionally, some studies \cite{r8,r9} have explored backdoor attacks that do not rely on training data, directly modifying model parameters during the training phase or the deployment phase.

\textbf{Multi-Target Backdoor Attacks.}
Research on multi-target backdoor attacks emphasizes broadening the backdoor payloads. Xue et al. propose two studies \cite{r11,r12} on broadening backdoor payloads in black-box settings. One \cite{r11} alters the strength of a patch trigger to serve as triggers for different target classes, and another \cite{r12} uses discrete cosine transform steganography in RGB channels to inject triggers targeting three classes. It demonstrates that utilizing RGB channels can effectively mitigate the similarity between different triggers. However, these studies have limitations in broadening backdoor payloads, can only target a few classes, and suffer from poor visual stealthiness.
Marksman \cite{r13} achieves powerful backdoor payloads in white-box settings by employing a class-conditional autoencoder trained along with the victim model.
It can target any class during inference, but the need for control over the training process and the limited transferability of generated triggers across models with significant structural differences restrict its applicability in black-box settings. Universal Backdoor Attacks \cite{r14} utilizes interclass poison transferability to achieve results similar to Marksman \cite{r13} in gray-box settings, that is, it requires precise knowledge of the victim model, which is not always practical.
Additionally, its triggers exhibit poor stealthiness due to the limitations of binary strings. Therefore, SFIBA is the first approach to effectively broaden the backdoor payloads capable of attacking all classes while ensuring trigger stealthiness in black-box settings.

\subsection{Backdoor Defenses}
Currently, the backdoor defense paradigms \cite{r24,r28,r25,r26} can be mainly categorized into input-based defenses, model-based defenses, output-based defenses, and inhibition-based defenses.
Input-based defenses directly detect or disrupt the trigger within samples, thereby rendering the backdoor ineffective. For example, frequency-based detector \cite{r34} identifies the trigger by detecting high-frequency artifacts.
Model-based defenses detect or eliminate backdoor by exploring or altering model parameters. For example, Gradcam \cite{r20} can easily observe the difference between poisoned and clean samples by plotting the model's attention in the form of a heatmap. Neural Cleanse \cite{r21} obtains possible triggers by reverse engineering each class of the model and comparing them to obtain the classes that are likely to be attacked. Fine-Pruning \cite{r22} destroies the backdoor structure by pruning the model. 
Output-based defenses detect the backdoor by validating model outputs.
For example, STRIP \cite{r23} perturbs potentially poisoned images with a random set of clean images and monitors the entropy of output to determine whether the model is infected with a backdoor. 
EBBA \cite{r33} calculates the energy of each class in model's output and looks for relatively anomalous high-energy values to determine the presence of backdoor.
Inhibition-based defenses aim to train clean models based on poisoned datasets. For example, CBD \cite{r24} learns backdoor-free models directly from contaminated datasets from a causal point of view. 
However, these defense mechanisms struggle to detect SFIBA.


\begin{figure*}[!t]
    \centering
    \includegraphics[width=0.95\textwidth]{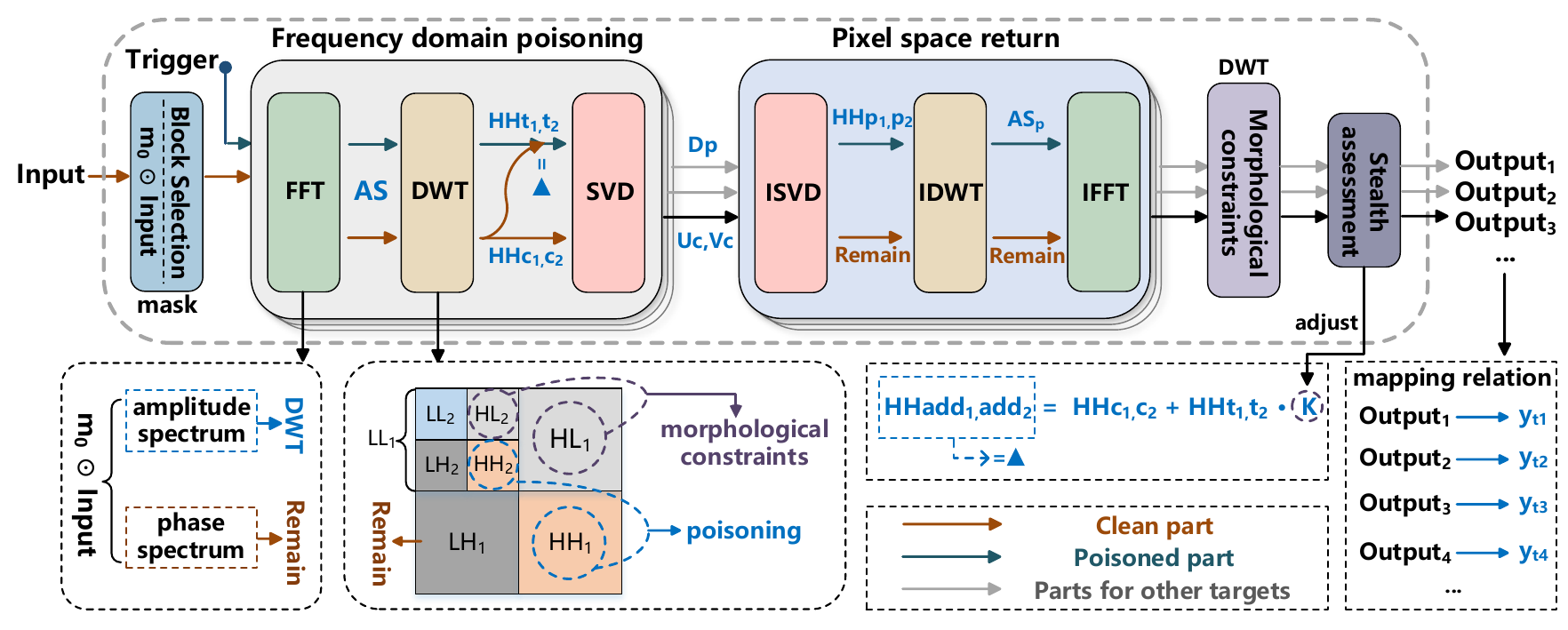}
    \vspace{-2mm}
    \caption{SFIBA's attack process, where $AS$ represents the amplitude spectrum, $HH_{c1,c2}$, $HH_{t1,t2}$, and $HH_{p1,p2}$ denote the diagonal features of the clean, trigger and poisoned  amplitude spectrum, respectively. }
    \label{fig1:env}
    \vspace{-6mm}
\end{figure*}

\section{Threat Model}
\label{sec:Threat_Model}
\subsection{Capability of Attackers}

SFIBA can perform full-target attacks in black-box settings. Specifically, the attacker only needs to manipulate the training set, without needing to know any knowledge of the model's architecture or parameters, nor interfere in the model training process. The attacker's permissions in above black-box settings are far more limited than those in previous backdoor attacks with powerful payloads. This significantly increases the threat posed by SFIBA in real-world attack scenarios.

\subsection{Attack Modeling}
In a typical image classification task, the goal is to train an accurate deep model $f : X \rightarrow C$, where $X$ represents the set of images and $C = \{c_1,c_2...,c_M\}$ denotes a set of $M$ classes.
After the SFIBA attack, the poisoned model $f'$ will classify poisoned samples $B_t(x)$ into corresponding predefined classes $y_t$, while maintaining performance on clean sample pairs $(x, y)$, as shown below:
\begin{equation}
\small
\begin{split}
&f'(x) = y,\quad  f'(B_t(x)) = y_t,\\
&x \in X,\quad  y, y_t \in C, \quad t \in \{1,2,...,M\},
\end{split}
\end{equation}
where $B_t(\cdot)$ represents the trigger injection paradigm specific to class $y_t$.
To realize above objective, the SFIBA attack process can be summarized as follows.
First, we randomly select a batch of clean samples and generate corresponding poisoned samples for each class, creating a poisoned set of size $N_p$, with $\lfloor N_p/M \rfloor$ samples per class. Then, we combine this poisoned set with a clean set of size $N_b$ for model training, as below:
\begin{equation}
\small
\underset{\theta}{\min}\sum_{i=1}^{N_b}\mathcal{L}(f'(x_i;\theta),y_i)+\sum_{t=1}^{M}\sum_{j=1}^{\lfloor N_p/M \rfloor}\mathcal{L}(f'(B_t(x_j);\theta),y_t),
\label{equation 4}
\end{equation}
where $\mathcal{L}$ stands for the cross-entropy loss.
The training process establishes a mapping between each trigger injection paradigm $B_t(\cdot)$ and its corresponding target class $y_t$, allowing flexible selection of attack targets during inference.

\section{Methodology}
\label{sec:Methodology}
\subsection{Motivation}
Previous experiments \cite{r27} have shown that backdoor models relying on static triggers are highly sensitive to both trigger spatial location and morphology. In the inference process, the Attack Success Rate (ASR) plummets with even minor shifts in the spatial location or changes in the morphology of static triggers. The backdoor's sensitivity to the trigger's spatial location is particularly significant for broadening backdoor payloads. To further elucidate this phenomenon, we build upon recent studies \cite{r30} on the Neural Tangent Kernel (NTK) to analyze backdoor behavior under variations in the trigger's spatial location. Specifically, we demonstrate that when the trigger is invisible, the backdoor effect is effectively eliminated if the trigger is shifted during the inference stage.
\begin{lemma}\label{thm:lemma1}
For the poisoned sample $ x_0' = (1 - m_0) \odot x + T(m_0 \odot x) $, the trigger is injected into the local region $ m_0 \odot x $, obtained by element-wise multiplication between the mask $ m_0 $ and the sample $ x $. The mask $ m_0 $ is a binary mask with a small rectangular region set to 1 in any position, while the rest of the elements are set to 0.
And the trigger injection paradigm $T(\cdot)$ is invisible, which has good visual indicators. The poisoned image classifier's output is $f'(x_0') = y_t$.
During inference, if the mask $m_0$ is altered to mask $m_1$ in such a way that the trigger position corresponding to $m_1$ no longer overlaps with that of $m_0$. For the poisoned sample $x_1' = (1-m_1) \odot x + T(m_1 \odot x)$, the probability that it is classified as $y_t$ is $\phi_t(x_1') < 0.5$. This indicates that $f'(x_1') \neq y_t$.
\end{lemma}
\IEEEpubidadjcol

\begin{figure}[!t]
\vspace{-2mm}
    \centering
    \begin{subfigure}{0.26\textwidth} 
        \centering
        \includegraphics[width=\textwidth]{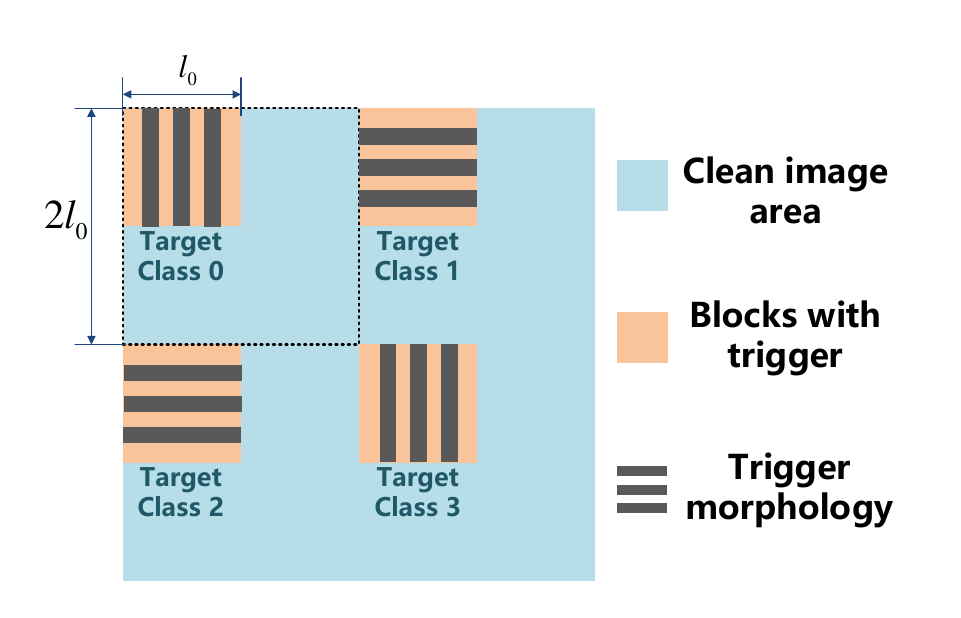}
        \caption{Distribution}
        \label{Blocks}
    \end{subfigure}
    \begin{subfigure}{0.19\textwidth} 
        \centering
        \includegraphics[width=\textwidth]{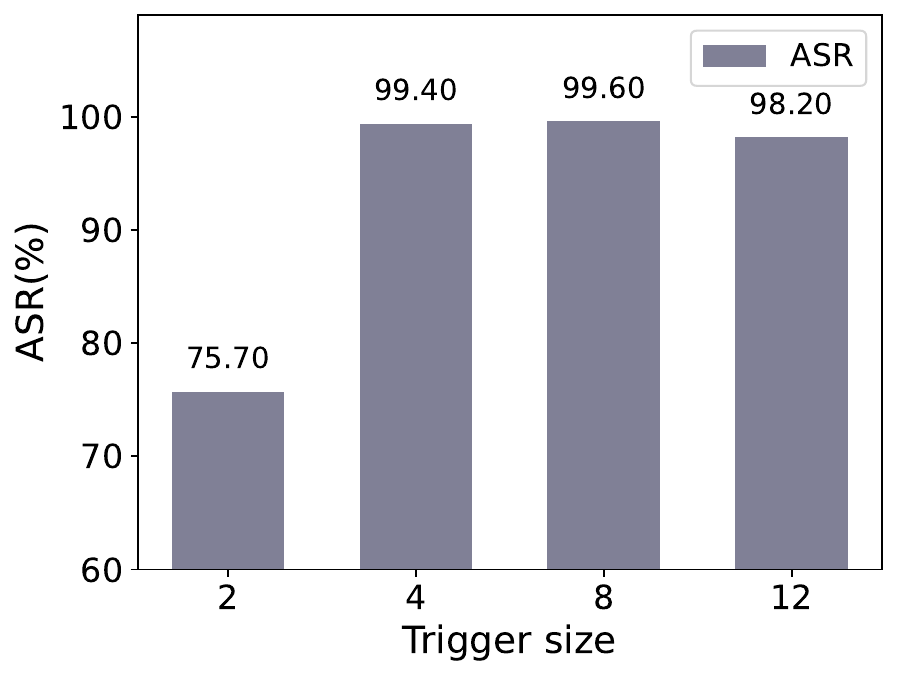}
        \caption{Effectiveness}
        \label{test_size}
    \end{subfigure}
    \caption{Distribution and effectiveness of $Block$s with triggers.}
    \label{fig:example2}
\vspace{-8mm}
\end{figure}


The proof of Lemma \ref{thm:lemma1} is provided in Appendix \ref{appendixA}. Here, we will further analyze Lemma 1, starting with a single-target attack using category $a$ as the target class, where the poisoned model is denoted as $f_a'$.
We theoretically prove that when the trigger is invisible, changing its position during inference prevents the poisoned sample from being classified into the predefined target class $a$. This indicates that the shifted trigger either does not activate or activates only a small number of backdoor neurons of class $a$. 
Thus, the neurons that $f_a'$ use to extract features from the shifted trigger and perform linear classification have an extremely low correlation with the backdoor neurons of class $a$. This means $f_a'$ is fully capable of adjusting the relevant neurons to establish a mapping between the shifted trigger and another target class $b$, thereby completing a dual-target backdoor attack.
By extension, we can achieve multi-target backdoor attacks through reasonable partitioning of the trigger's spatial positions.
In summary, \textbf{Lemma 1 indicates that as long as injecting invisible triggers into non-overlapping local regions to attack different target classes, we can minimize mutual interference among the various backdoors during training, thereby achieving multi-target backdoor attacks.}

Building on this foundational theory, we propose restricting triggers to specific local spatial regions (i.e., $Block$s) and applying different morphological constraints based on their location to ensure the specificity of triggers for different target classes.
Specifically, we divide the image into mutually isolated blocks in the pixel space, with each block denoted as $Block_i$, where $i$ serves as the serial number of $Block$ for identification. For each target class $y_t$, we systematically select a $Block_i$ and one of its RGB channels for trigger injection as:
\begin{equation}
\small
\begin{split}
&B_t(x) = x - Block_i + T_{n,i}(Block_i),\\
&Block_i = m_i \odot x,\quad S = \lfloor M/3 \rfloor,\\
&i = t \bmod S,  \quad n = \lfloor t/S \rfloor, \quad t \in \{1,2,...,M\}, 
\end{split}
\end{equation}
where $B_t(x)$ represents the obtained poisoned sample and $t$ indicates the target class is $y_t$. $Block_i$ denotes the trigger injection position determined by the mask $m_i$. 
$T_{n,i}(\cdot)$ represents the injection of trigger with specific morphological constraints. Here, $i$ determines the type of morphological constraint and $n$ determines the channel for trigger injection. Particularly, when $n$ is 0, 1, and 2, triggers are injected into the B, R, and G channels, respectively.
This means that once all $Block$s in a channel are exhausted, we switch to the next channel for trigger injection. 
\textbf{It should be noted that each poisoned image uses only one channel of a single Block for trigger injection, uniquely corresponding to a specific target $y_t$.} 
Furthermore, we introduce an invisible dynamic trigger injection paradigm to maintain the stealthiness and effectiveness of triggers within the finite $Block$s. 
In summary, as shown in Fig. \ref{fig1:env}, the generation of poisoned images involves three key steps: block selection, frequency domain poisoning, and trigger morphology constraints along with dynamic optimization. Each of these steps will be elaborated on in the following sections.

\subsection{Local Space Dynamic Invisible Trigger Injection}
\textbf{Step 1: Block Selection.}
We define the length and width of a clean sample as $(L,W)$ respectively and divide $Block_i$ into a square, with a side length of $l_0$. The number of target classes is $M$. Consider that Data Aaugmentation (DA) such as translation, rotation, flipping, and clipping during the training phase, can alter the spatial location of $Block_i$. 
This may lead to overlap between $Block_i$ and $Block_{i'} (i' \neq i)$ in different poisoned samples, potentially compromising the effectiveness of the attack. To address this issue, we opt to add an interval around $Block_i$ so that:
\begin{equation}
\small
\begin{split}
Distance(Block_i, Block_{around}) = 2l_0, \\ L, W \bmod 2l_0 \geq l_0, \quad i \in \{1,2,...,\lfloor M / 3)\rfloor\},
\end{split}
\end{equation}
where $Distance(Block_i, Block_{around})$ represents the minimum distance between $Block_i$ and other $Block$s around it.
Thus, each $Block_i$ actually occupies a square region of $2l_0$ on each side and the trigger is injected into the second quadrant of this region, as illustrated in Fig. \ref{Blocks}. 
This strategy reduces the likelihood of multiple triggers overlapping due to data augmentation.
With this setup, we develop a class-specific single-channel $Block$ selection algorithm, as shown in Algorithm \ref{alg1}.
Using this algorithm, we can obtain the spatial location of the unique single-channel $Block$ corresponding to each target class.
In addition, we apply distinct morphological constraints to the triggers in adjacent $Block_i$, further enhancing the trigger specificity. The above methods significantly improve the robustness of SFIBA against data augmentation, as will be detailed in the experimental section.

\renewcommand{\thealgorithm}{1}
\begin{algorithm}
\caption{Class-specific $Block$ Fetching Algorithm}
\label{alg1}
\begin{algorithmic}[1]
\Require the length and width of clean sample $L, W$, the length of $Block$ $l_0$;
\Ensure $Block$ list; \\
\textbf{Initialize: }$n_L=\lfloor L/2l_0\rfloor$, $n_W=\lfloor W/2l_0\rfloor$, Target label list $=\{0,1,2,3,4,5...\}$, $Block$ list = \{\};  
\If{$n_L \cdot n_W \geq \lfloor M/3 \rfloor$}  \hfill\Comment{single channel $Block$ num}

\While{$num$ in Target label list}
\State $n_1=\lfloor num/n_W \rfloor \bmod n_W$; \hfill \Comment{the row of $Block$}
\State $n_2=num \bmod n_L$; \hfill \Comment{the column of $Block$}
\State $Block_{num} = \text{clean image}[n_1 \cdot 2l_0 : n_1 \cdot 2l_0+l_0; n_2 \cdot 2l_0 : n_2 \cdot 2l_0+l_0]$; 
\State Add $Block_{num}$ to $Block$ list;
\EndWhile
\EndIf
\State \Return $Block$ list
\end{algorithmic}
\end{algorithm}

\textbf{Step 2: Frequency Domain Poisoning.}
Injecting a trigger into $Block$ in the frequency domain involves three key steps: frequency-domain transformation using FFT, feature extraction from the amplitude spectrum via DWT, and trigger injection through singular value fusion. The specific operations and functions of each step are as follows:

\textbf{(1) Frequency-domain transformation using FFT.}
We introduce FFT to initially ensure the trigger stealthiness.
FFT can decompose an image into amplitude and phase spectra. 
According to studies \cite{r17,r29}, the amplitude spectrum captures low-level semantic information, while the phase spectrum captures high-level semantics. Modifying the amplitude spectrum is unlikely to significantly alter high-level semantics and visual appearance. 
Therefore, we extract the amplitude and phase spectra of the clean $Block$, as well as the amplitude spectrum of the trigger image. 
To ensure trigger stealthiness, we retain the clean phase spectrum and attempt to inject the trigger's amplitude spectrum information into the clean amplitude spectrum.
However, conventional methods of injecting triggers into the amplitude spectrum, such as directly overlaying the trigger spectrum on the central region of the clean spectrum, are not suitable for small $Block$.
There are two main reasons:

\begin{itemize}[itemindent=0pt, left=0pt]
\item \textbf{Difficulty in selecting the injection region: } Overlaying the trigger across the entire amplitude spectrum would significantly compromise stealthiness. Therefore, triggers are usually superimposed on a fixed local region of the amplitude spectrum. For complete images, frequency energy is typically concentrated in the low-frequency region, making it easier to select injection region. However, for small $Blocks$, the unpredictable content leads to irregular energy distributions, complicating the selection of a uniform injection region that balances trigger effectiveness and stealthiness.
\item \textbf{Difficulty in selecting the injection coefficient: } Each Small $Block$ occupies an extremely small area. For example, each $Block$ from the ImageNet100 dataset covers only 0.22\% of the total image area (as detailed in the hyperparameters section). Therefore, fine-tuning the injection coefficient will significantly impact the stealthiness and effectiveness of the trigger, making it difficult to decide.
\end{itemize}

In summary, we exploit the semantic properties of the amplitude spectrum obtained by FFT to initially ensure trigger stealthiness. Ablation studies show that, while maintaining visual quality, removing the FFT reduces SFIBA's average ASR on CIFAR10 to 10.1\%. This underscores the necessity of FFT. However, due to the limitations of small $Block$, relying solely on FFT for trigger injection is not feasible.

\textbf{(2) Amplitude spectrum feature extraction via DWT.}
To address the challenge of selecting the trigger injection region, we extract features from the entire amplitude spectrum of $Block$ and inject the trigger into specific features. \textbf{In particular, we treat the amplitude spectrum as an image $x_{as}$ in the pixel space and use DWT to extract its features as follows:}
\begin{equation}
\small
\begin{split}
&DWT(x_{as}) = \{LL_1, HL_1, LH_1, HH_1\},\\
&DWT(LL_1) = \{LL_2, HL_2, LH_2, HH_2\},
\end{split}
\end{equation}
where $LL_1$, $HL_1$, $LH_1$, and $HH_1$ denote the approximation subgraphs, longitudinal edge features, horizontal edge features, and diagonal features, respectively. Among these, the approximation subgraphs contain the majority of the energy, while the diagonal features contain relatively little energy. In addition, we can recursively apply the same feature extraction process to $LL_1$. 
It is evident that DWT can effectively extract features in different directions from the entire amplitude spectrum. 
We multiply the diagonal features of trigger's amplitude spectrum by an injection coefficient $K$ and superimpose them onto the clean diagonal components $HH1$ and $HH2$, as shown below:
\begin{equation}
\small
\begin{aligned}
HH_{add1,add2}=HH_{c1,c2}+HH_{t1,t2} \times K, 
\end{aligned}
\end{equation}
 in which $HH_{c1},HH_{c2}$ are the diagonal features of clean amplitude spectrum, $HH_{t1},HH_{t2}$ are those of trigger's amplitude spectrum. Injecting trigger into both high-frequency $HH_1$ and low-frequency $HH_2$ components enhances its robustness and effectiveness. Additionally, the low energy in the diagonal features enhances trigger stealthiness. 
 
In summary, DWT solves the challenge of selecting the trigger injection region and enhances both trigger stealthiness and effectiveness. Its necessity will be further illustrated in the ablation study section.
In addition, it is crucial to note that directly applying DWT to $Block$ in the pixel space for trigger injection is infeasible. This is because DWT extracts features directly from the spatial domain, and superimposing triggers on these features can easily alter critical ones, degrading visual quality. To maintain stealthiness, the trigger's intensity must be reduced, which would cause the attack to fail. Thus, neither FFT nor DWT alone can achieve an invisible attack under the constraints of $Block$. Only their combination can ensure both trigger stealthiness and effectiveness.

 \textbf{(3) Trigger injection through singular value fusion.}
 To address the challenge of selecting the trigger injection coefficient, we reduce the sensitivity of the trigger strength to changes in the injection coefficient, thereby enhancing its adjustability. Specifically, we introduce SVD to convert the direct superposition of the trigger in the amplitude spectrum into an indirect variation of singular values, as follows:
\begin{equation}
\small
\begin{aligned}
&SVD(HH_{m})=U_{m}D_{m}V^T_{m},\quad m = \{add1,add2,c1,c2\},\\
&HH_{p1,p2}=ISVD(U_{c1,c2}D_{add1,add2}V^T_{c1,c2}),
\label{EQ9}
\end{aligned}
\end{equation}
where $D_m$ represents the singular value matrix, ISVD denotes the inverse process of SVD and $HH_{p1},HH_{p2}$ represent the final poisoned diagonal features of the amplitude spectrum. The singular values in $D_m$ capture the stable embedded properties of an image and tend to remain consistent even under minor perturbations. Thus, retaining only the poisoned information in the singular values reduces the sensitivity of the trigger strength to the injection coefficient and simplifies the selection of the optimal injection coefficient.

Subsequently, we combine $HH_{p1},HH_{p2}$ with the remaining clean features and perform the inverse DWT (IDWT) to generate the poisoned amplitude spectrum as Eq. (\ref{equation 9}). Here, $AS_{p}$ denotes the resulting poisoned amplitude spectrum. 
\begin{equation}
\small
\begin{split}
&LL_{p1} = IDWT\{LL_{c2}, HL_{c2}, LH_{c2}, HH_{p2}\},\\
&AS_{p} = IDWT\{LL_{p1}, HL_{c1}, LH_{c1}, HH_{p1}\}.\\
\end{split}
\label{equation 9}
\end{equation}
Finally, we utilize $AS_{p}$ along with the clean phase spectrum to generate poisoned $Block$ via the inverse FFT.

\textbf{Step 3: Trigger Morphology Constraints along with Dynamic Optimization.}
To enhance trigger specificity and robustness against data augmentation, we impose differentiated morphological constraints on triggers in adjacent $Blocks$.
Specifically, we perform DWT on both poisoned $Block$ and original clean $Block$, retaining trigger information only in either horizontal or vertical features, while replacing the rest with clean features.
Taking horizontal features as an example, we preserve only the poisoned high- and low-frequency horizontal features $LH_{p1},LH_{p2}$, as follows:
\begin{equation}
\small
\begin{split}
&DWT(Block_{poisoned}) \Rightarrow \{LH_{p}\}, \\
&DWT(Block_{clean}) \Rightarrow \{LL_{c}, HL_{c}, HH_{c}\},\\
&Block_{horizontal} = IDWT\{LL_{c}, HL_{c}, LH_{p}, HH_{c}\},\\ &c = \{c1,c2\},p = \{p1,p2\}.
\end{split}
\end{equation}
This results in the trigger being horizontally distributed. In contrast, triggers in adjacent $Blocks$ are restricted to vertical distribution. 
To verify the effectiveness of the DWT-based trigger morphology constraints, we conducted validation experiments. 
We used two types of triggers in the same $Block$ and channel but with different morphologies to attack two categories.
The ASR for the two targets are 90.8\% and 89.2\%, respectively. This demonstrates that DWT-based trigger constraints effectively enhance trigger specificity.
Ultimately, we reintegrate the modified $Block$ back into the original image, completing the construction of the poisoned image.  \textbf{Each complete poisoned sample contains just a single trigger within one channel of a Block.}


\renewcommand{\thealgorithm}{2}
\begin{algorithm}
\caption{PSNR-based Trigger Dynamic Tuning Algorithm}
\label{alg2}
\begin{algorithmic}[1]
\Require PSNR thresholds $(p_0,p_1)$, range of injection coefficient $(K_{min},K_{max})$, maximum Iterations $n_0$, clean sample $image_{c}$, current poisoned sample $image_{p}$;
\Ensure optimal injection coefficient $K$;\\
\textbf{Initialize: } $n=0$, $left, right=K_{min,max}$, $K=K_{min}$;
\While{$n<n_0$}
\State $PSNR=psnr(image_{c},image_{p})$;
\If{$K=K_{min}$}
    \If{$PSNR<p_0$ or $PSNR$ in $(p_0,p_1)$}
        \State\Return {$K$} \hfill \Comment{early stopping}
    \Else:
        \State $K=K_{max}$; \hfill \Comment{prepare to use dichotomy}
    \EndIf
\Else:
    \If{$PSNR$ in $(p_0,p_1)$}
        \State\Return {$K$} \hfill \Comment{coefficient that meets conditions}
    \ElsIf{$PSNR<p_0$}
        \State $right=K$; \hfill \Comment{adjust right boundary}
        \State $K=(left+right)/2$; 
    \Else:
        \State $left=K$; \hfill \Comment{adjust left boundary}
        \State $K=(left+right)/2$;
    \EndIf
\EndIf
\State update $image_{p}$;
\State $n=n+1$;
\EndWhile
\State\Return {$K$} \hfill \Comment{coefficient at the end of the iteration}
\end{algorithmic}
\end{algorithm}

Finally, to ensure high stealthiness for each poisoned sample, we dynamically adjust the trigger injection coefficient $K$ based on visual metrics.
Commonly used visual metrics include Peak Signal-to-Noise Ratio (PSNR), Structural Similarity Index (SSIM), and Learned Perceptual Image Patch Similarity (LPIPS). However, frequent computation of all these metrics would significantly slow down the poisoned sample generation process. Therefore, we propose a dynamic adjustment algorithm based solely on PSNR as shown in in Algorithm \ref{alg2}.
Specifically, we set the PSNR thresholds as ($p_1$ and $p_0$) and the range of $K$ as $(K_{min}, K_{max})$. To ensure trigger effectiveness, we maintain $K \geq K_{min}$ even if the PSNR falls below $p_0$. In the context of ensuring effectiveness, we accept the poisoned samples only when their PSNR fall within the range $(p_0, p_1)$ to ensure trigger stealthiness; otherwise, we adjust $K$ using the dichotomy method.
It is worth noting that the spatial and morphological constraints of the trigger ensure its specificity remains robust, even when the trigger strengths are similar.



In summary, SFIBA embeds dynamic, invisible, and effective triggers in specific $Blocks$, leveraging the sensitivity of the backdoor to trigger location and morphology, thus effectively broadening the backdoor payloads in black-box settings.

\begin{table}[!t]
\setlength{\tabcolsep}{4.5pt} 
\small
\centering
\caption{Attack properties comparison.}
\vspace{-1mm}
\label{properties-table}
\begin{tabular}{ccccc}
\toprule
\multirow{3}{*}{Methods} & \multicolumn{4}{c}{Properties (\ding{51}/\ding{55})} \\ 
\cmidrule{2-5}
                         & \begin{tabular}[c]{@{}c@{}}Visual \\ Stealthiness\end{tabular} & \begin{tabular}[c]{@{}c@{}}Full \\ Target\end{tabular} & \begin{tabular}[c]{@{}c@{}}Black-Box \\ Settings\end{tabular} & \begin{tabular}[c]{@{}c@{}}Benign \\ Accuracy\end{tabular}\\ 
\midrule
SFIBA                     & \ding{51} & \ding{51} & \ding{51} & \ding{51} \\ 
One-to-N \cite{r11}                 & \ding{55} & \ding{55} & \ding{51} & \ding{55} \\ 
Marksman \cite{r13}                & \ding{55} & \ding{51} & \ding{55} & \ding{51} \\ 
UBA \cite{r14}                     & \ding{55} & \ding{51} & \ding{55} & \ding{51} \\ 
\bottomrule
\end{tabular}

\vspace{-5mm}
\end{table}

\section{Evaluation}
\label{sec:Evaluation}
\begin{table*}[!t]
\centering
  \caption{Average attack performance of SFIBA and baselines (PRN denoting PreActResNet).}
  \label{sample-table1}
  \normalsize 
\begin{tabular}{*{8}{p{2cm}}}
    \toprule
    \multicolumn{1}{c|}{} &\multicolumn{1}{c|}{} &\multicolumn{2}{c|}{\centering CIFAR10} & \multicolumn{2}{c|}{\centering GTSRB} & \multicolumn{2}{c}{\centering ImageNet100} \\
    \multicolumn{1}{c|}{\multirow{-2}{1.5cm}{\centering Method}}&\multicolumn{1}{c|}{\multirow{-2}{1.5cm}{\centering Indicators}} &\multicolumn{1}{c|}{\centering PRN} & \multicolumn{1}{c|}{\centering VGG} &\multicolumn{1}{c|}{\centering PRN} & \multicolumn{1}{c|}{\centering VGG} &\multicolumn{1}{c|}{\centering ResNet} & \multicolumn{1}{c}{\centering VGG} \\
    
    \cmidrule(r){1-1}\cmidrule(r){2-2} \cmidrule(r){3-4} \cmidrule(r){5-6}\cmidrule(r){7-8}
    
    \multicolumn{1}{c|}{} &\multicolumn{1}{c|}{ASR}    & \multicolumn{1}{c|}{0.6078} & \multicolumn{1}{c|}{0.5842}  & \multicolumn{1}{c|}{0.1402} & \multicolumn{1}{c|}{0.1523}  & \multicolumn{1}{c|}{0.1164} & \multicolumn{1}{c}{0.1013} \\
    \multicolumn{1}{c|}{} &\multicolumn{1}{c|}{BA}     & \multicolumn{1}{c|}{0.9114} & \multicolumn{1}{c|}{0.8817}  & \multicolumn{1}{c|}{0.9677} & \multicolumn{1}{c|}{0.9621}  & \multicolumn{1}{c|}{0.8222} & \multicolumn{1}{c}{0.8344} \\
    \multicolumn{1}{c|}{\multirow{-3}{3cm}{\centering One-to-N \cite{r11}}}&\multicolumn{1}{c|}{DV}     & \multicolumn{1}{c|}{0.0342} & \multicolumn{1}{c|}{0.0262}  & \multicolumn{1}{c|}{0.0121} & \multicolumn{1}{c|}{0.0103}  & \multicolumn{1}{c|}{0.0606} & \multicolumn{1}{c}{0.0650} \\
    \cmidrule(r){1-1}\cmidrule(r){2-2} \cmidrule(r){3-4} \cmidrule(r){5-6}\cmidrule(r){7-8}
    
    \multicolumn{1}{c|}{} &\multicolumn{1}{c|}{ASR}    & \multicolumn{1}{c|}{0.9989} & \multicolumn{1}{c|}{1.0000}  & \multicolumn{1}{c|}{0.9998} & \multicolumn{1}{c|}{1.0000}  & \multicolumn{1}{c|}{0.9994} & \multicolumn{1}{c}{0.9948} \\
    \multicolumn{1}{c|}{} &\multicolumn{1}{c|}{BA}     & \multicolumn{1}{c|}{0.9320} & \multicolumn{1}{c|}{0.9078}  & \multicolumn{1}{c|}{0.9789} & \multicolumn{1}{c|}{0.9712}  & \multicolumn{1}{c|}{0.8108} & \multicolumn{1}{c}{0.8568} \\
    \multicolumn{1}{c|}{\multirow{-3}{3cm}{\centering Marksman \cite{r13}}}&\multicolumn{1}{c|}{DV}     & \multicolumn{1}{c|}{0.0136} & \multicolumn{1}{c|}{0.0001}  & \multicolumn{1}{c|}{0.0009} & \multicolumn{1}{c|}{0.0012}  & \multicolumn{1}{c|}{0.0720} & \multicolumn{1}{c}{0.0426} \\
    \cmidrule(r){1-1}\cmidrule(r){2-2} \cmidrule(r){3-4} \cmidrule(r){5-6}\cmidrule(r){7-8}
    
    \multicolumn{1}{c|}{} &\multicolumn{1}{c|}{ASR}    & \multicolumn{1}{c|}{0.9445} & \multicolumn{1}{c|}{1.0000}  & \multicolumn{1}{c|}{0.9975} & \multicolumn{1}{c|}{0.9961}  & \multicolumn{1}{c|}{0.9980} & \multicolumn{1}{c}{0.9990} \\
    \multicolumn{1}{c|}{} &\multicolumn{1}{c|}{BA}     & \multicolumn{1}{c|}{0.8675} & \multicolumn{1}{c|}{0.8523}  & \multicolumn{1}{c|}{0.9555} & \multicolumn{1}{c|}{0.9633}  & \multicolumn{1}{c|}{0.8755} & \multicolumn{1}{c}{0.8828} \\
    \multicolumn{1}{c|}{\multirow{-3}{3cm}{\centering UBA \cite{r14} \\ without DA }}&\multicolumn{1}{c|}{DV}     & \multicolumn{1}{c|}{0.0100} & \multicolumn{1}{c|}{0.0136}  & \multicolumn{1}{c|}{0.0243} & \multicolumn{1}{c|}{0.0081}  & \multicolumn{1}{c|}{0.0005} & \multicolumn{1}{c}{0.0202} \\

    \cmidrule(r){1-1}\cmidrule(r){2-2} \cmidrule(r){3-4} \cmidrule(r){5-6}\cmidrule(r){7-8}

    \multicolumn{1}{c|}{} &\multicolumn{1}{c|}{ASR}    & \multicolumn{1}{c|}{0.9972} & \multicolumn{1}{c|}{0.9954}  & \multicolumn{1}{c|}{0.9975} & \multicolumn{1}{c|}{0.9994}  & \multicolumn{1}{c|}{0.9924} & \multicolumn{1}{c}{0.9928} \\
    \multicolumn{1}{c|}{} &\multicolumn{1}{c|}{BA}     & \multicolumn{1}{c|}{0.9368} & \multicolumn{1}{c|}{0.9020}  & \multicolumn{1}{c|}{0.9755} & \multicolumn{1}{c|}{0.9721}  & \multicolumn{1}{c|}{0.8674} & \multicolumn{1}{c}{0.8924} \\
    \multicolumn{1}{c|}{\multirow{-3}{1.5cm}{\centering SFIBA with DA }}&\multicolumn{1}{c|}{DV}     & \multicolumn{1}{c|}{0.0088} & \multicolumn{1}{c|}{0.0059}  & \multicolumn{1}{c|}{0.0043} & \multicolumn{1}{c|}{0.0003}  & \multicolumn{1}{c|}{0.0154} & \multicolumn{1}{c}{0.0070} \\
    \cmidrule(r){1-1}\cmidrule(r){2-2} \cmidrule(r){3-4} \cmidrule(r){5-6}\cmidrule(r){7-8}
    \multicolumn{1}{c|}{} &\multicolumn{1}{c|}{ASR}    & \multicolumn{1}{c|}{0.9984} & \multicolumn{1}{c|}{0.9958}  & \multicolumn{1}{c|}{0.9964} & \multicolumn{1}{c|}{0.9992}  & \multicolumn{1}{c|}{0.9958} & \multicolumn{1}{c}{0.9933} \\
    \multicolumn{1}{c|}{} &\multicolumn{1}{c|}{BA}     & \multicolumn{1}{c|}{0.8774} & \multicolumn{1}{c|}{0.8645}  & \multicolumn{1}{c|}{0.9738} & \multicolumn{1}{c|}{0.9702}  & \multicolumn{1}{c|}{0.8662} & \multicolumn{1}{c}{0.8948} \\
    \multicolumn{1}{c|}{\multirow{-3}{2cm}{\centering SFIBA without DA }}&\multicolumn{1}{c|}{DV}     & \multicolumn{1}{c|}{0.0001} & \multicolumn{1}{c|}{0.0014}  & \multicolumn{1}{c|}{0.0060} & \multicolumn{1}{c|}{0.0012}  & \multicolumn{1}{c|}{0.0098} & \multicolumn{1}{c}{0.0082} \\
    
    \bottomrule
\end{tabular}
\end{table*}

\begin{table*}[!t]
\centering
\caption{ASR of SFIBA for each class in CIFAR10.}
\label{sample-table3}
\normalsize 
\begin{tabular}{*{12}{p{2cm}}}
\toprule
\multicolumn{1}{c|}{} & \multicolumn{1}{c|}{DA} & \multicolumn{10}{c}{Target class} \\ 
\cmidrule(r){2-2} \cmidrule(r){3-12} 
\multicolumn{1}{c|}{\multirow{-2}{1cm}{\centering Model}} & \multicolumn{1}{c|}{(\ding{51} / \ding{55} )} & \multicolumn{1}{c|}{1} & \multicolumn{1}{c|}{2} & \multicolumn{1}{c|}{3} & \multicolumn{1}{c|}{4} & \multicolumn{1}{c|}{5} & \multicolumn{1}{c|}{6} & \multicolumn{1}{c|}{7} & \multicolumn{1}{c|}{8} & \multicolumn{1}{c|}{9} & \multicolumn{1}{c}{10} \\ 
\cmidrule(r){1-1} \cmidrule(r){2-2} \cmidrule(r){3-12} 
\multicolumn{1}{c|}{} & \multicolumn{1}{c|}{\ding{51}} & \multicolumn{1}{c|}{0.998} & \multicolumn{1}{c|}{1.000} & \multicolumn{1}{c|}{0.990} & \multicolumn{1}{c|}{1.000} & \multicolumn{1}{c|}{1.000} & \multicolumn{1}{c|}{0.996} & \multicolumn{1}{c|}{0.996} & \multicolumn{1}{c|}{0.994} & \multicolumn{1}{c|}{1.000} & \multicolumn{1}{c}{0.998} \\ 
\multicolumn{1}{c|}{\multirow{-2}{1cm}{\centering PRN}} & \multicolumn{1}{c|}{\ding{55}} & \multicolumn{1}{c|}{0.998} & \multicolumn{1}{c|}{1.000} & \multicolumn{1}{c|}{0.998} & \multicolumn{1}{c|}{0.996} & \multicolumn{1}{c|}{1.000} & \multicolumn{1}{c|}{0.998} & \multicolumn{1}{c|}{0.994} & \multicolumn{1}{c|}{1.000} & \multicolumn{1}{c|}{1.000} & \multicolumn{1}{c}{1.000} \\ 
\cmidrule(r){1-1} \cmidrule(r){2-2} \cmidrule(r){3-12} 
\multicolumn{1}{c|}{} & \multicolumn{1}{c|}{\ding{51}} & \multicolumn{1}{c|}{0.994} & \multicolumn{1}{c|}{0.996} & \multicolumn{1}{c|}{0.992} & \multicolumn{1}{c|}{0.996} & \multicolumn{1}{c|}{0.996} & \multicolumn{1}{c|}{0.998} & \multicolumn{1}{c|}{0.992} & \multicolumn{1}{c|}{0.990} & \multicolumn{1}{c|}{1.000} & \multicolumn{1}{c}{1.000} \\ 
\multicolumn{1}{c|}{\multirow{-2}{1cm}{\centering VGG}} & \multicolumn{1}{c|}{\ding{55}} & \multicolumn{1}{c|}{0.998} & \multicolumn{1}{c|}{0.996} & \multicolumn{1}{c|}{0.994} & \multicolumn{1}{c|}{0.996} & \multicolumn{1}{c|}{0.998} & \multicolumn{1}{c|}{0.996} & \multicolumn{1}{c|}{0.990} & \multicolumn{1}{c|}{0.990} & \multicolumn{1}{c|}{1.000} & \multicolumn{1}{c}{1.000} \\ 
\bottomrule
\end{tabular}
\vspace{-4mm}
\end{table*}

\subsection{Experimental Setup}
\textbf{Baselines.}
We utilize three most advanced multi-target backdoor attacks, One-to-N \cite{r11}, Marksman \cite{r13}, and Universal Backdoor Attacks (UBA) \cite{r14} as our baselines. 
One-to-N \cite{r11} is one of the most representative black-box multi-target attack paradigms. 
Marksman \cite{r13} and UBA\cite{r14} implement the most powerful backdoor payloads in non-black-box settings.
For an accurate assessment of multiple baseline methods, we adhere strictly to the prerequisites for attack privileges they demand. TABLE \ref{properties-table} presents the properties of each attack paradigm. It is evident that SFIBA is the only approach to achieve full-target attacks while ensuring trigger stealthiness in black-box settings.

\textbf{Datasets and Models.}
We assess the effectiveness of the SFIBA using several datasets commonly utilized for backdoor attacks, including CIFAR10, GTSRB, and ImageNet100. For CIFAR10 and GTSRB, we employ PreActResnet18 and VGG19 models, while for ImageNet100, we use pre-trained versions of Resnet18 and VGG19 models. Additionally, for the baselines, we apply the same datasets and models to maintain consistency in our evaluation.

\textbf{Hyperparameters.}
For each model's training, we employ the momentum SGD optimizer with an initial learning rate of 0.001 and a momentum of 0.9. The learning rate is decayed by a factor of 0.1 every 30 epochs. 
During the training process, we adopt the same data augmentation method as WANET \cite{r5}, which involves various spatial transformation techniques capable of effectively changing the spatial location of the trigger.
When generating poisoned samples, we set the PSNR limits $(p_0,p_1)$ to $(40,42)$ and the injection coefficient $K$ tuning range to $(0.1,40)$. 
Additionally, we tested the attack effectiveness  with different-sized $Block$ using single-target attacks on CIFAR10, and the results are shown in Fig \ref{test_size}. It can be seen that the attack is effective when the absolute size of the $Block$ is at least $4 \times 4$. 
Consequently, we set the $Block$ size and complete poisoned sample size for different datasets as follows: For CIFAR10, the $Block$ size is $8 \times 8$, with a complete sample size of $32 \times 32$, allowing for 12 attack targets. For GTSRB, the $Block$ size is $8 \times 8$, with a complete sample size of $64 \times 64$, enabling 48 attack targets. For ImageNet100, the $Block$ size is $12 \times 12$, and the complete sample size is $256 \times 256$, accommodating up to 300 attack targets.
Thus, for each dataset, we can easily achieve the full-target attack. We use 3\% of the data to generate poisoned samples for CIFAR10 and GTSRB, and 2.5\% for ImageNet100. For each baseline, we replicate its optimal experimental setup to ensure comparable and successful results. 

\begin{figure*}[!t]
    \centering
    \includegraphics[width=0.7\textwidth]{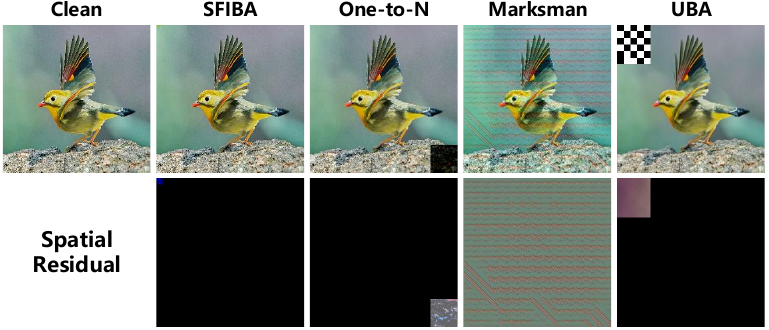}
    \vspace{-1mm}
    \caption{Visual effects and residuals of SFIBA and baselines on ImageNet. }
    \label{picture_SFIBA_baseline}
    \vspace{-3mm}
\end{figure*}

\subsection{Effectiveness of SFIBA}

\begin{table}[!t]
\normalsize
  \centering
  \caption{SFIBA's visual metrics across various datasets.}
  \label{table5}
  \begin{tabular}{*{4}{p{2cm}}}
    \toprule
    \multicolumn{1}{c|}{Dataset}    & \multicolumn{1}{c|}{PSNR} & \multicolumn{1}{c|}{SSIM}  & \multicolumn{1}{c}{LPIPS}\\
    \midrule
    \multicolumn{1}{c|}{CIFAR10}    & \multicolumn{1}{c|}{40.752} & \multicolumn{1}{c|}{0.9951}  & \multicolumn{1}{c}{0.0001} \\
    \multicolumn{1}{c|}{GTSRB}     & \multicolumn{1}{c|}{40.545} & \multicolumn{1}{c|}{0.9934}  & \multicolumn{1}{c}{0.0113}\\
    \multicolumn{1}{c|}{ImageNet100}     & \multicolumn{1}{c|}{40.250} & \multicolumn{1}{c|}{0.9975}  & \multicolumn{1}{c}{0.0095}\\
    \bottomrule
  \end{tabular}
\vspace{-5mm}
\end{table}

\textbf{Attack Effectiveness.}
TABLE \ref{sample-table1} shows SFIBA's attack performance with and without DA, compared to the baselines. The DA methods, akin to WANET \cite{r5}, involve various spatial transformation techniques. We report the average ASR across all classes, the Benign classification Accuracy (BA), and the Decrease Value in BA (DV) compared to the clean model. Whether with or without DA, SFIBA achieves a high ASR across all classes and datasets, with negligible impact on BA. Therefore, it is evident that our design of spatial intervals and morphological constraints on the triggers effectively resists DA. 
SFIBA demonstrates significant advantages over various baselines. It outperforms One-to-N in ASR across all three datasets and has a smaller impact on BA, clearly surpassing the most representative black-box multi-target attack method. Compared to Marksman, SFIBA achieves similar ASR and DV but requires significantly fewer permissions, making it a greater practical threat. Against Universal Backdoor Attacks, SFIBA excels on CIFAR10 and performs similarly on other datasets. The limitations of Universal Backdoor Attacks lie in the constraints of binary code length and the requirement for large patch sizes, making them difficult to apply to datasets with fewer categories and small-sized images. In contrast, SFIBA offers stable attack performance across various scenarios, supporting any number of categories and a range of image sizes.
TABLE \ref{sample-table3} demonstrates SFIBA's strong performance across all CIFAR10 classes, with DA having a negligible effect on its attack efficacy. This underscores SFIBA's robustness against trigger location shifts caused by DA.

\textbf{Attack Stealthiness.}
We assess the stealthiness of poisoned samples generated by SFIBA using the PSNR, SSIM, and LPIPS metrics. 
Specifically, we randomly choose 50 clean samples from each dataset. For each clean sample, we create multiple poisoned samples targeting all categories and calculate their average visual metrics. As shown in TABLE \ref{table5}, these poisoned samples demonstrate outstanding visual metrics.
In addition, Fig. \ref{picture_SFIBA_baseline} compares the poisoned sample generated by SFIBA on ImageNet100 with the one generated by each baseline. It shows that SFIBA achieves excellent visual quality, significantly outperforming the baselines. 
\vspace{-3mm}

\subsection{Performance Against Defensive Measures}

We evaluate the robustness of SFIBA against popular backdoor defense mechanisms including Fine-Pruning \cite{r22}, Neural Cleanse \cite{r21}, CBD \cite{r24}, STRIP \cite{r23} and EBBA \cite{r33}. These backdoor defense mechanisms have achieved remarkable results in defending against previous backdoor attacks.

\begin{figure*}[!t]
    \centering
    \begin{subfigure}{0.32\textwidth} 
        \centering
        \includegraphics[width=\textwidth]{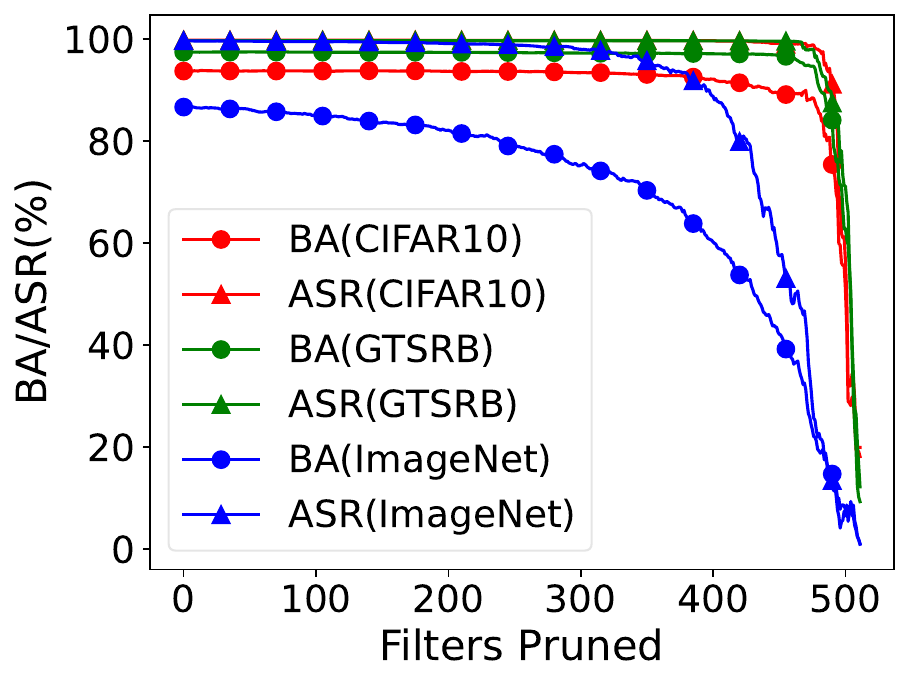}
        \caption{Fine-Pruning}
        \label{Fine-Pruning}
    \end{subfigure}
    \begin{subfigure}{0.32\textwidth} 
        \centering
        \includegraphics[width=\textwidth]{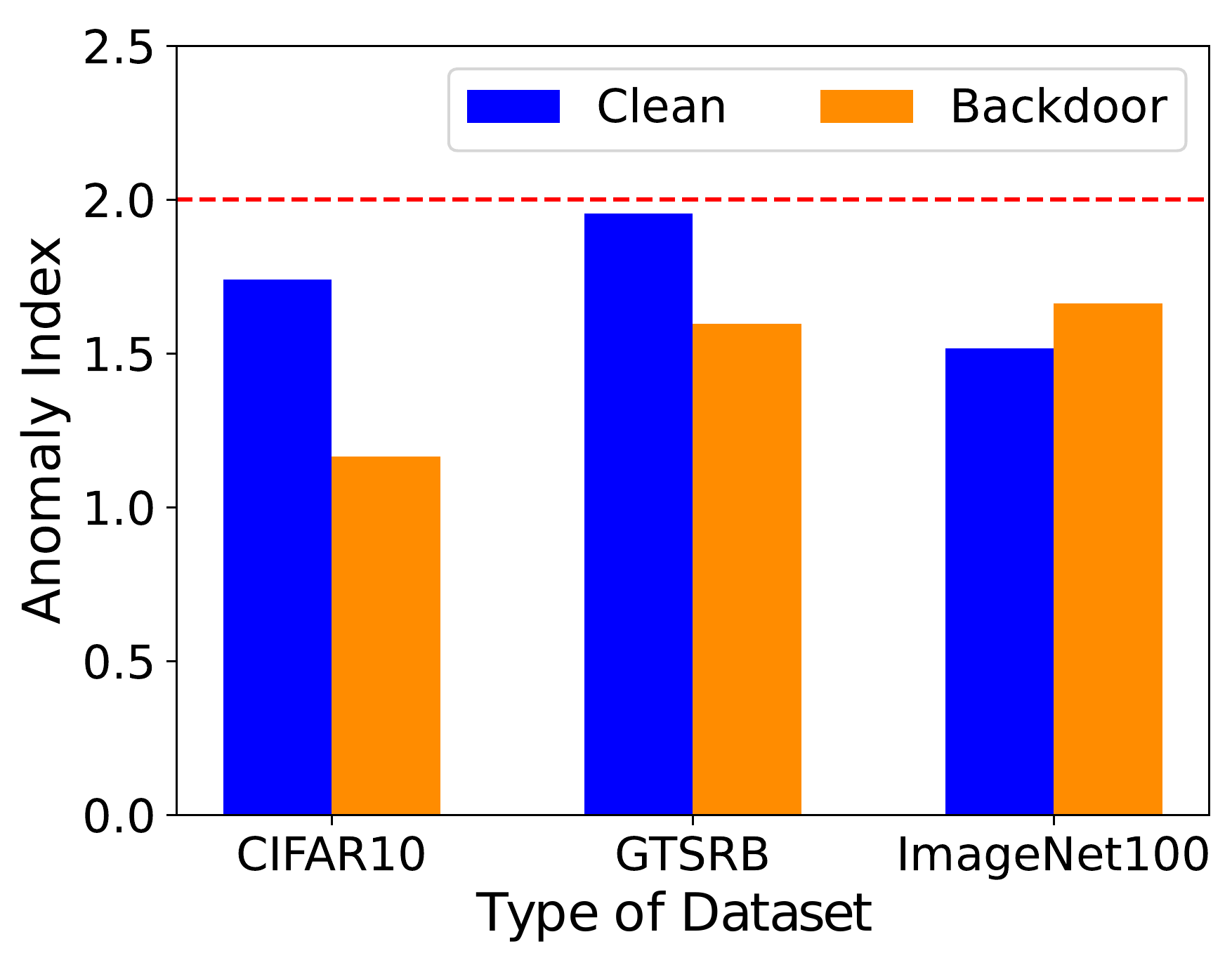}
        \caption{Neural Cleanse}
        \label{Neural Cleanse}
    \end{subfigure}
    \begin{subfigure}{0.32\textwidth} 
        \centering
        \includegraphics[width=\textwidth]{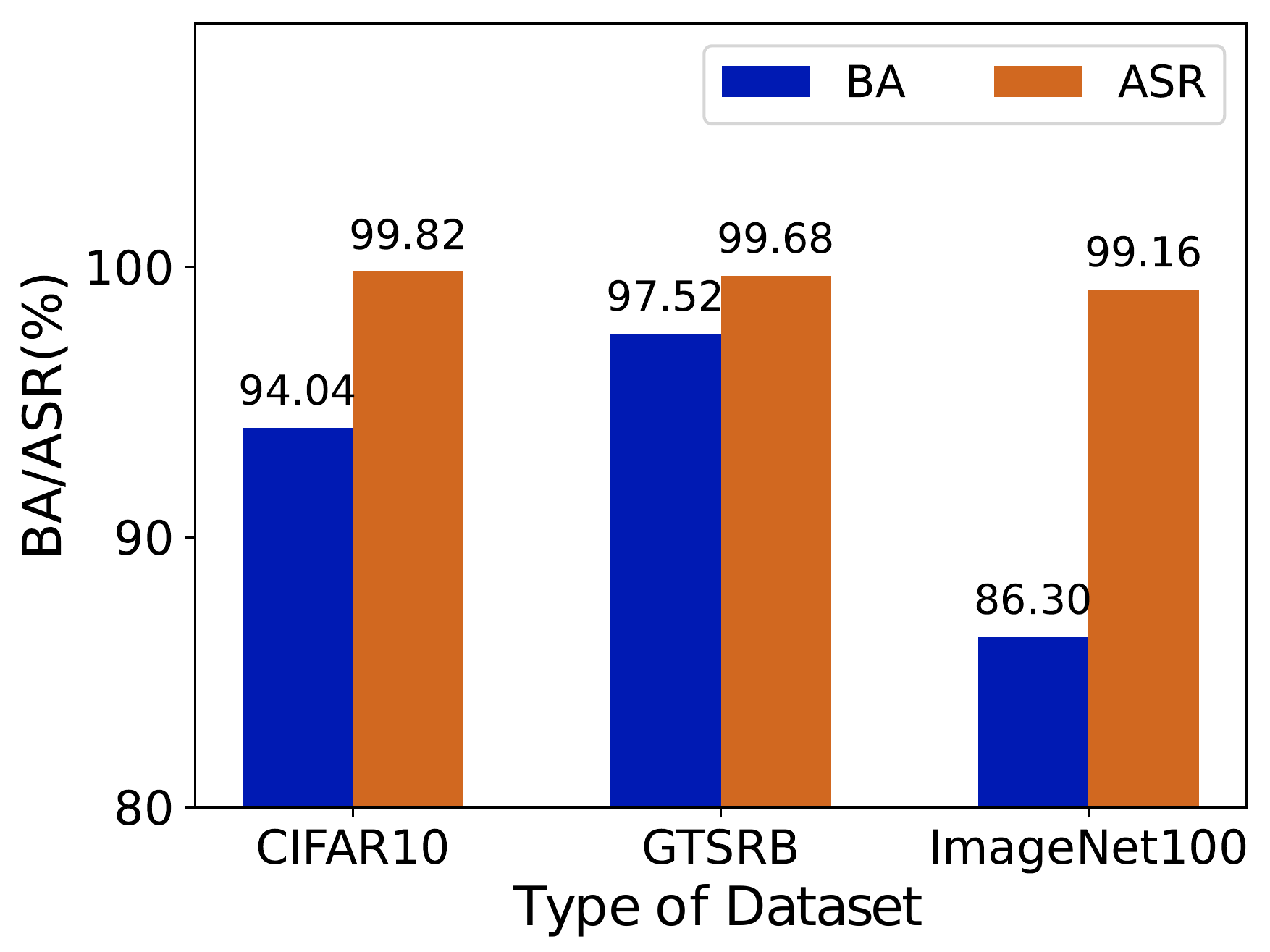}
        \caption{CBD}
        \label{CBD}
    \end{subfigure}
    \vspace{-1mm}
    \caption{The performance of SFIBA in resisting Fine-Pruning, Neural Cleanse, and CBD.}
    \label{all_defense}
\vspace{-3mm}
\end{figure*}

\begin{figure*}[!t]
    \centering
    \begin{subfigure}{0.32\textwidth} 
        \centering
        \includegraphics[width=\textwidth]{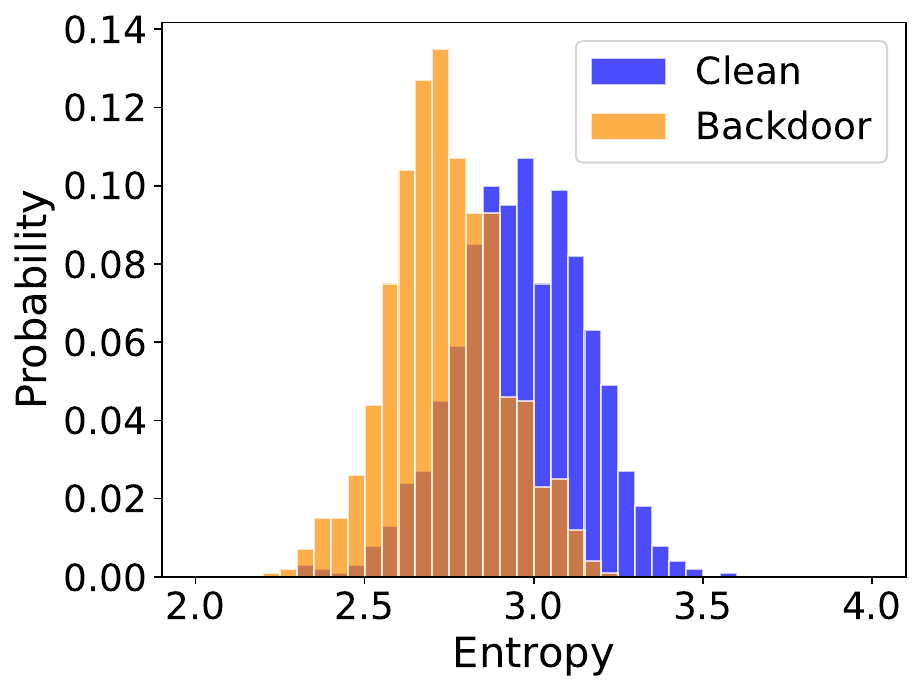}
        \caption{CIFAR10}
        \label{CIFAR10}
    \end{subfigure}
    \begin{subfigure}{0.32\textwidth} 
        \centering
        \includegraphics[width=\textwidth]{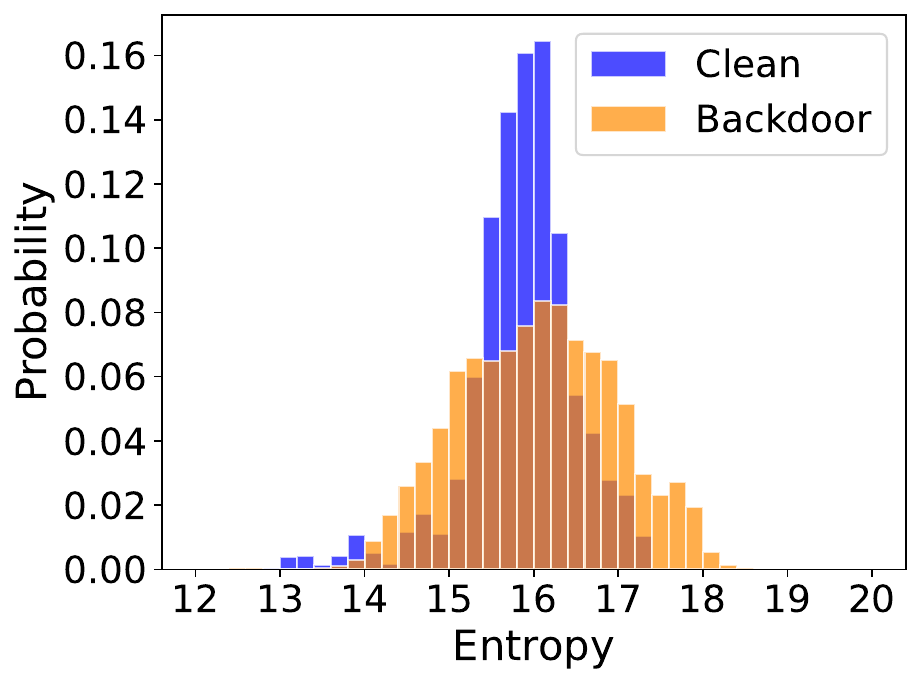}
        \caption{GTSRB}
        \label{GTSRB}
    \end{subfigure}
    \begin{subfigure}{0.32\textwidth} 
        \centering
        \includegraphics[width=\textwidth]{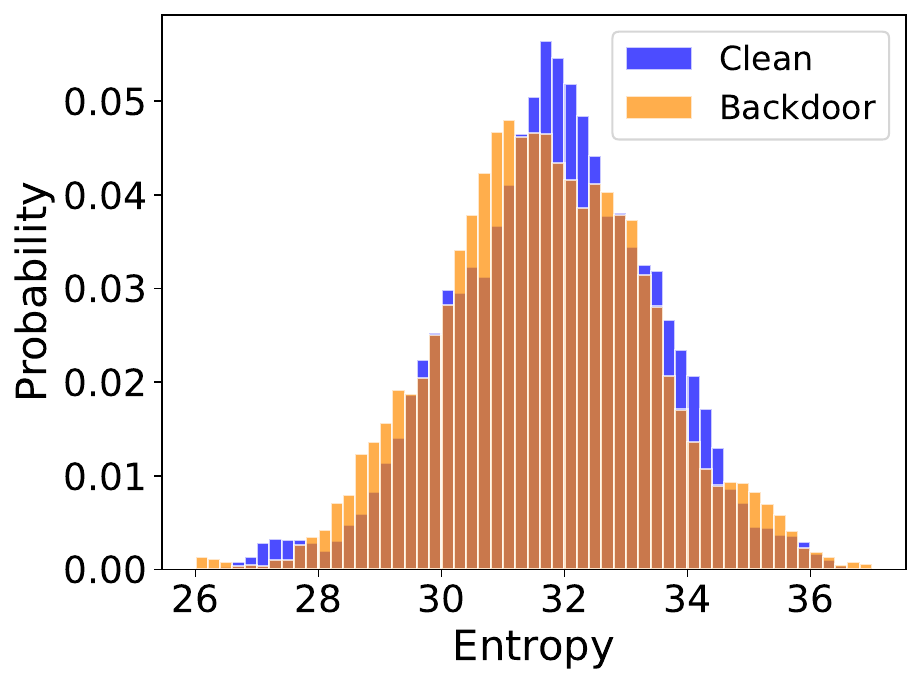}
        \caption{ImageNet100}
        \label{ImageNet100}
    \end{subfigure}
    \vspace{-1mm}
    \caption{The performance of SFIBA in resisting STRIP.}
    \label{all_defense_strip}
\vspace{-3mm}
\end{figure*}

\begin{figure*}[!t]
    \centering
    \begin{subfigure}{0.32\textwidth} 
        \centering
        \includegraphics[width=\textwidth]{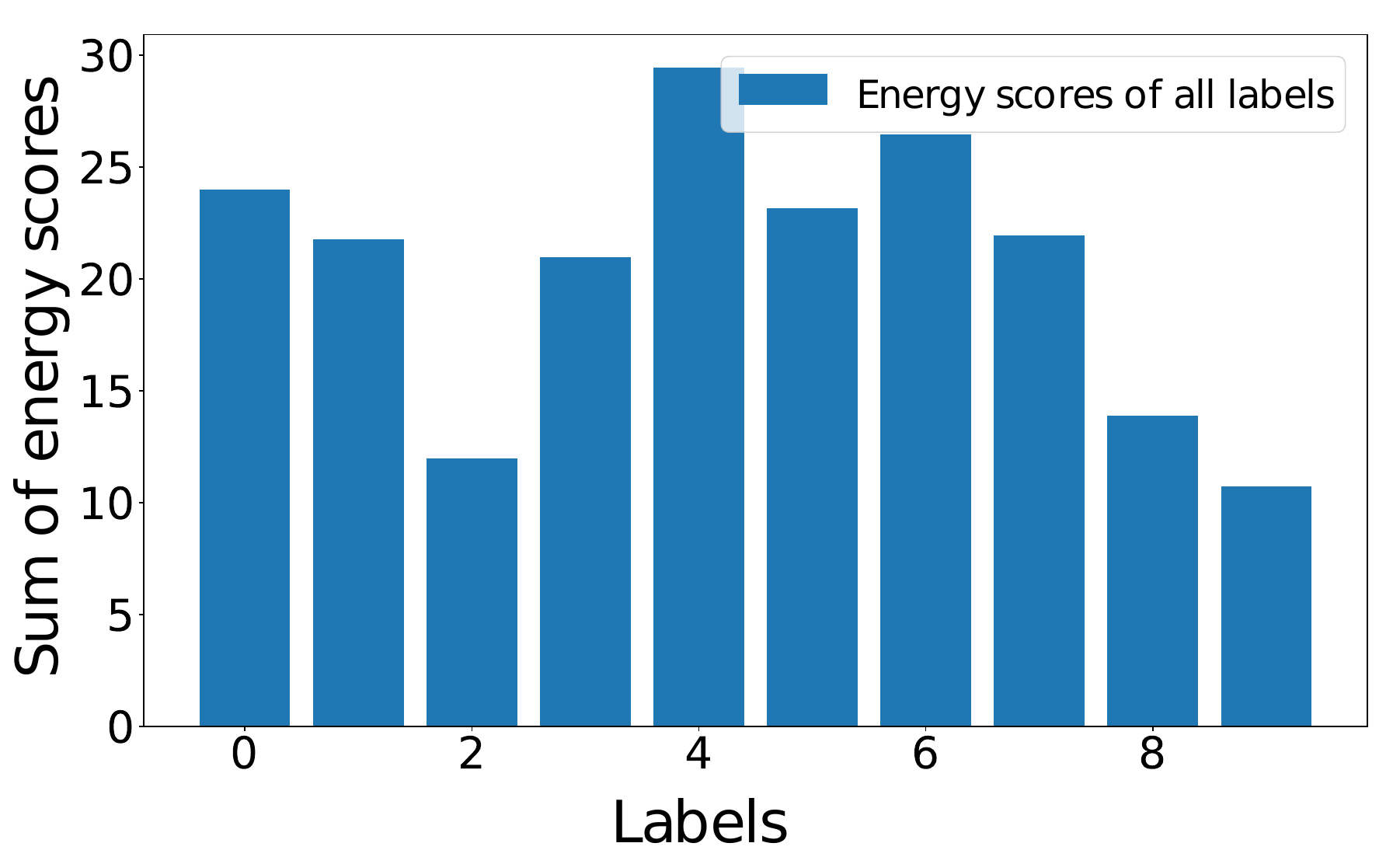}
        \caption{CIFAR10}
        \label{EBBA_CIFAR10}
    \end{subfigure}
    \begin{subfigure}{0.32\textwidth} 
        \centering
        \includegraphics[width=\textwidth]{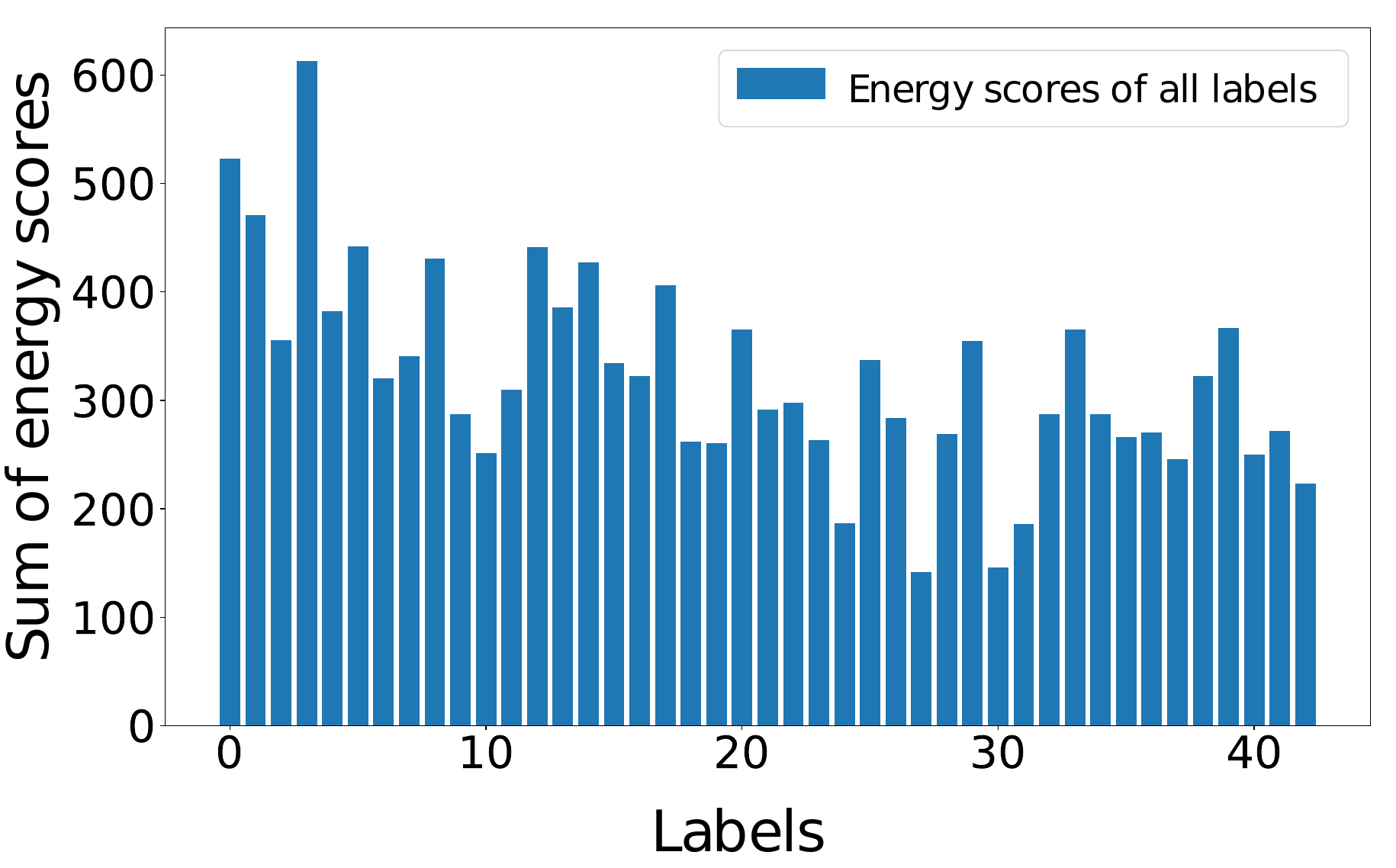}
        \caption{GTSRB}
        \label{EBBA_GTSRB}
    \end{subfigure}
    \begin{subfigure}{0.32\textwidth} 
        \centering
        \includegraphics[width=\textwidth]{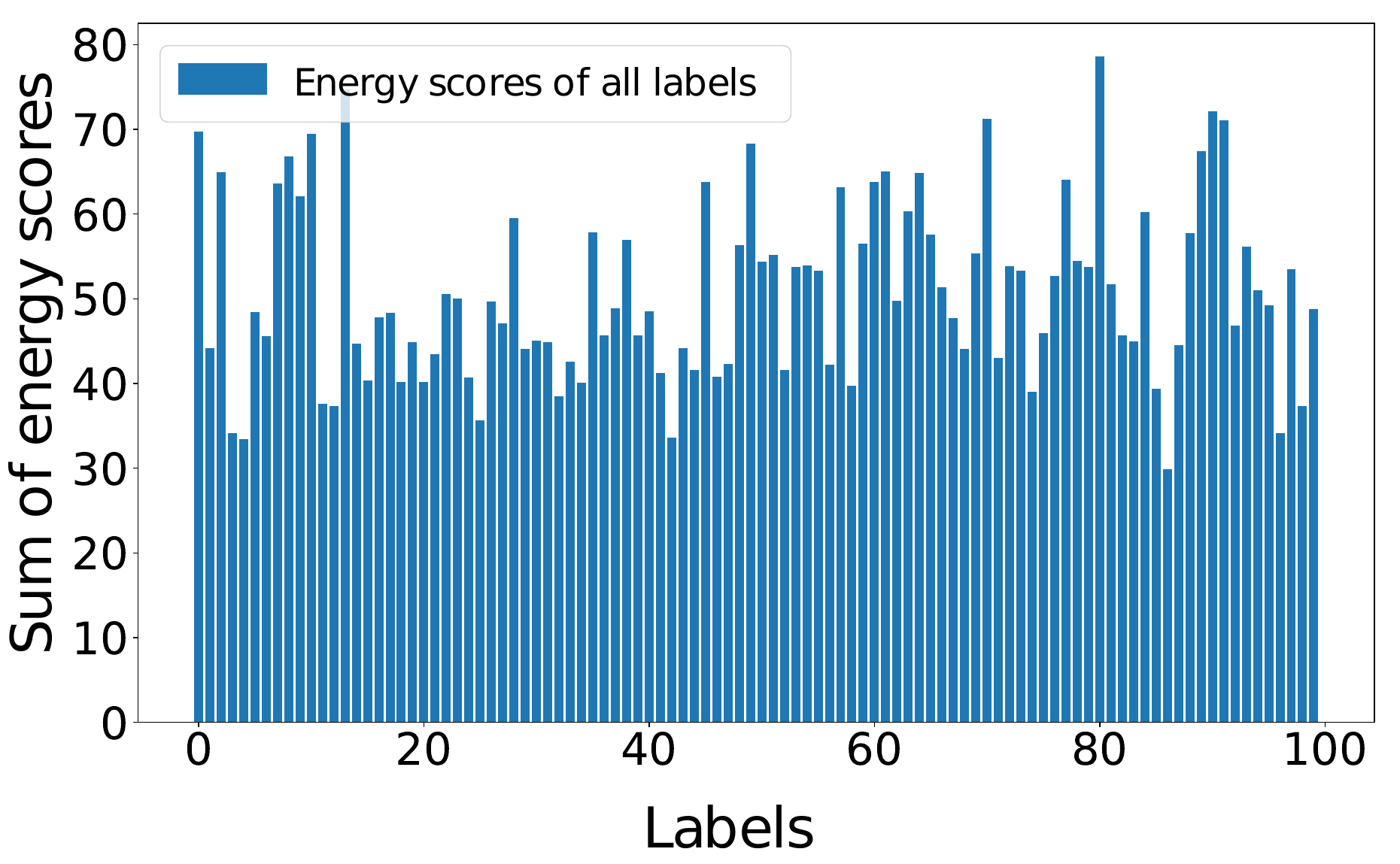}
        \caption{ImageNet100}
        \label{EBBA_ImageNet100}
    \end{subfigure}
    \vspace{-1mm}
    \caption{The performance of SFIBA in resisting EBBA.}
    \label{all_defense_EBBA}
\vspace{-5mm}
\end{figure*}
\textbf{Resistance to Fine-Pruning.}
We assess SFIBA's robustness to Fine-Pruning, which prunes dormant neurons for backdoor removal while maintaining benign accuracy. Fig. \ref{Fine-Pruning} shows the impact of neuron pruning on both BA and average ASR. The ASR reduction is consistently less than the decline in BA, indicating SFIBA's strong stealthiness against Fine-Pruning.

\textbf{Resistance to Neural Cleanse.}
We evaluate SFIBA's robustness against Neural Cleanse, which uses inverse triggers to detect backdoors with an anomaly metric above 2. Figure \ref{Neural Cleanse} shows that the anomaly metrics for both benign models and poisoned models are below 2, indicating that SFIBA exhibits high stealthiness against Neural Cleanse.

\textbf{Resistance to CBD.}
We then assess SFIBA's robustness against CBD, which mitigates the confounding effect by learning causality and using a sample-by-sample weighting scheme on a contaminated dataset. Fig. \ref{CBD} presents the results when training with CBD. SFIBA demonstrates high ASR and BA, indicating its robustness to CBD.

\textbf{Resistance to STRIP.}
We test SFIBA's robustness to STRIP, which perturbs clean images to create high-entropy outputs in benign models. Low entropy indicates a backdoor. Fig. \ref{all_defense_strip} shows similar entropy distributions for clean and poisoned samples on CIFAR10, GRSRB and ImageNet100, suggesting SFIBA's strong stealthiness against STRIP.

\textbf{Resistance to EBBA.}
Finally, we test SFIBA's robustness against EBBA, which calculates the energy of each class in the model's output. If any class shows a significantly high energy value, it suggests the presence of a backdoor. We compute the energy for each class across three datasets, with results presented in Fig. \ref{all_defense_EBBA}. Since SFIBA treats all classes as targets, EBBA cannot identify any relatively high anomalous energy values. Therefore, EBBA cannot detect SFIBA.


\begin{table}[!t]
\normalsize
  \centering
  \caption{SFIBA's visual metrics during ablation process.}
  \label{Step1}
  \begin{tabular}{*{4}{p{2cm}}}
    \toprule
    \multicolumn{1}{c|}{Step}    & \multicolumn{1}{c|}{PSNR} & \multicolumn{1}{c|}{SSIM}  & \multicolumn{1}{c}{LPIPS}\\
    \midrule
    \multicolumn{1}{c|}{Step 1}    & \multicolumn{1}{c|}{40.579} & \multicolumn{1}{c|}{0.9932}  & \multicolumn{1}{c}{0.0019} \\
    \multicolumn{1}{c|}{Step 2}    & \multicolumn{1}{c|}{40.785} & \multicolumn{1}{c|}{0.9944}  & \multicolumn{1}{c}{0.0024} \\
    \multicolumn{1}{c|}{Step 3}    & \multicolumn{1}{c|}{40.886} & \multicolumn{1}{c|}{0.9945}  & \multicolumn{1}{c}{0.0023} \\
    \multicolumn{1}{c|}{Step 4}     & \multicolumn{1}{c|}{40.333} & \multicolumn{1}{c|}{0.9914}  & \multicolumn{1}{c}{0.0019} \\
    \bottomrule
  \end{tabular}
\vspace{-5mm}
\end{table}

\begin{table*}[!t]
\normalsize
\centering
  \caption{Attack performance for each class in each step.}
  \label{Step1_1}
  \begin{tabular}{*{12}{p{2cm}}}
\toprule
    \multicolumn{1}{c|}{} &\multicolumn{1}{c|}{} &\multicolumn{10}{c}{ASR for each target class label}\\
    \cmidrule(r){3-12} 
    \multicolumn{1}{c|}{\multirow{-2}{1cm}{\centering Step}}  & 
    \multicolumn{1}{c|}{\multirow{-2}{1cm}{\centering BA}}  &\multicolumn{1}{c|}{1} & \multicolumn{1}{c|}{2}  & \multicolumn{1}{c|}{3} & \multicolumn{1}{c|}{4}  & \multicolumn{1}{c|}{5} & \multicolumn{1}{c|}{6}  & \multicolumn{1}{c|}{7} & \multicolumn{1}{c|}{8}  & \multicolumn{1}{c|}{9} & \multicolumn{1}{c}{10}\\
    \cmidrule(r){1-1}\cmidrule(r){2-2}\cmidrule(r){3-12} 
    \multicolumn{1}{c|}{\centering Step 1}    &\multicolumn{1}{c|}{0.936}    & \multicolumn{1}{c|}{1.000} & \multicolumn{1}{c|}{0.994}  & \multicolumn{1}{c|}{0.992} & \multicolumn{1}{c|}{1.000}  & \multicolumn{1}{c|}{0.996} & \multicolumn{1}{c|}{1.000}  & \multicolumn{1}{c|}{1.000} & \multicolumn{1}{c|}{0.996}  & \multicolumn{1}{c|}{0.996} & \multicolumn{1}{c}{0.998} \\
    \multicolumn{1}{c|}{\centering Step 2}    &\multicolumn{1}{c|}{0.932}    & \multicolumn{1}{c|}{0.978} & \multicolumn{1}{c|}{0.990}  & \multicolumn{1}{c|}{0.984} & \multicolumn{1}{c|}{0.982}  & \multicolumn{1}{c|}{0.984} & \multicolumn{1}{c|}{0.996}  & \multicolumn{1}{c|}{0.980} & \multicolumn{1}{c|}{0.974}  & \multicolumn{1}{c|}{0.988} & \multicolumn{1}{c}{0.986} \\
    \multicolumn{1}{c|}{\centering Step 3}    &\multicolumn{1}{c|}{0.931}    & \multicolumn{1}{c|}{0.970} & \multicolumn{1}{c|}{0.984}  & \multicolumn{1}{c|}{0.982} & \multicolumn{1}{c|}{0.978}  & \multicolumn{1}{c|}{0.982} & \multicolumn{1}{c|}{0.994}  & \multicolumn{1}{c|}{0.984} & \multicolumn{1}{c|}{0.978}  & \multicolumn{1}{c|}{0.986} & \multicolumn{1}{c}{0.978} \\
    \multicolumn{1}{c|}{\centering Step 4}    &\multicolumn{1}{c|}{0.907}    & \multicolumn{1}{c|}{0.212} & \multicolumn{1}{c|}{0.260}  & \multicolumn{1}{c|}{0.326} & \multicolumn{1}{c|}{0.260}  & \multicolumn{1}{c|}{0.468} & \multicolumn{1}{c|}{0.482}  & \multicolumn{1}{c|}{0.382} & \multicolumn{1}{c|}{0.408}  & \multicolumn{1}{c|}{0.364} & \multicolumn{1}{c}{0.372} \\
    \bottomrule
  \end{tabular}
\vspace{-5mm}
\end{table*}

\vspace{-3mm}

\subsection{Ablation Experiment of SFIBA}
To validate the necessity of each step in SFIBA, we conducted comprehensive ablation experiments on the PreActResNet18 model using the CIFAR10 dataset. The results indicate that each module of SFIBA is essential.

\textbf{Step 1: Remove Dynamic Optimization.}
Initially, we eliminate the dynamic adjustment of the trigger injection coefficient $K$, and subsequently select an optimal $K$ manually for the remaining portion, ultimately settling on a value of 2.5. We then randomly generate a set of poisoned samples and assess their visual effects, as presented in Step 1 of TABLE \ref{Step1}. 
While the average visual quality of the poisoned samples remains commendable at this stage, it is evident that there has been a notable decline in performance as indicated by both the SSIM and LPIPS metrics.
Step 1 in TABLE \ref{Step1_1} presents the attack performance for each class at this stage. The performance remains impressive, indicating that the remaining components are highly effective. However, this does not imply that dynamic adjustment is futile. We have already observed a decline in visual metrics when dynamic adjustment is removed on smaller datasets. For larger datasets, it would be even more challenging to manually select an appropriate injection coefficients $K$. 
Therefore, the stealthiness achieved with a dynamic injection coefficients $K$ is undoubtedly superior to that of a static one.

\textbf{Step 2: Remove Trigger Morphology Constraints.}
We remove the DWT used for trigger morphology constraints and manually select an optimal K for the remaining part, finally determining K to be 0.25. The average visual effects of the poisoned images at this stage, as shown in Step 2 of TABLE \ref{Step1}, still achieve a level similar to those before. Based on this, we test the attack performance at this stage, as shown in Step 2 of TABLE \ref{Step1_1}. It can be seen that the ASR at this point is no longer satisfactory, showing a significant decline. Therefore, the trigger morphology constraints is very important.

\textbf{Step 3: Remove Singular Value Fusion.}
We then eliminate singular value fusion in the frequency domain poisoning stage and directly overlay the diagonal features of the amplitude spectrum to inject the trigger. Subsequently, we manually set the injection coefficient $K$ to 0.24 to achieve the desired visual effects of the poisoned images, as shown in Step 3 of TABLE \ref{Step1}. Step 3 of TABLE \ref{Step1_1} presents the attack performance at this stage, revealing a significant decrease in ASR across multiple targets. This clearly demonstrates that singular value fusion can enhance the trigger's effectiveness and adjustability.

\textbf{Step 4: Remove DWT Feature Extraction.}
Finally, we eliminate the DWT component for amplitude spectrum feature extraction and directly superimpose trigger information on the amplitude spectrum of the $Block$. 
After adjusting the superposition coefficient to 0.05, the poisoned samples can achieve good average visual effects, as shown in Step 4 of TABLE \ref{Step1}. Step 4 of TABLE \ref{Step1_1} shows the current attack results, indicating a significant decrease in ASR for all targets, along with a certain reduction in BA. This is consistent with our previous analyses: without DWT-based amplitude spectrum feature extraction, the injection coefficient $K$ must be significantly lowered to maintain the stealthiness of the poisoned images, which substantially reduces the effectiveness of the trigger.

In summary, detailed ablation experiments demonstrate that each module in SFIBA is crucial, significantly contributing to the stealthiness and effectiveness of the triggers.

\section{Conclusion}
\label{sec:Conclusion}
In this paper, we introduce a novel backdoor attack called SFIBA, which can simultaneously attack all classes in black-box settings, that is, constructing class-specific trigger injection methods and establishing mappings between these methods and their corresponding target classes. This enables the classification of arbitrary poisoned samples into any target class during inference, while preserving the model's performance on benign samples. Moreover, SFIBA exhibits excellent visual stealthiness.
Specifically, SFIBA draws on the basic theory that backdoors are sensitive to both the spatial locations and morphologies of triggers and injects invisible, specific, and effective triggers into limited local regions using frequency-domain methods. 
Our experimental results on various datasets validate SFIBA's effectiveness, visual stealthiness, and robustness against existing defense mechanisms.
\appendix  
\subsection{Proof of Lemma 1}  
\label{appendixA}
\textbf{Lemma 1.}
\textit{
For the poisoned sample $ x_0' = (1 - m_0) \odot x + T(m_0 \odot x) $, the trigger is injected into the local region $ m_0 \odot x $, obtained by element-wise multiplication between the mask $ m_0 $ and the sample $ x $. The mask $ m_0 $ is a binary mask with a small rectangular region set to 1 in any position, while the rest of the elements are set to 0.
And the trigger injection paradigm $T(\cdot)$ is invisible, which has good visual indicators. The poisoned image classifier's output is $f'(x_0') = y_t$.
During inference, if the mask $m_0$ is altered to mask $m_1$ in such a way that the trigger position corresponding to $m_1$ no longer overlaps with that of $m_0$. For the poisoned sample $x_1' = (1-m_1) \odot x + T(m_1 \odot x)$, the probability that it is classified as $y_t$ is $\phi_t(x_1') < 0.5$. This indicates that $f'(x_1') \neq y_t$.
}

\begin{proof}
Consider a poisoned training dataset comprising $N_b$ benign samples and $N_P$ poisoned samples. The $N_b$ benign samples are obtained by uniformly sampling from $M$ classes of samples. Suppose that DNN $f'(\cdot;\theta)$ is a multivariate kernel regression (RBF kernel), trained on the poisoned training dataset using cross-entropy loss.
Following theory \cite{r31,r32}, we can get the regression solution for NTK as:
\begin{equation}
\begin{split}
\small
\phi_t(\cdot) = \frac{\sum_{i=1}^{N_b} K(\cdot,x_i) \cdot y_i + \sum_{i=1}^{N_p} K(\cdot,x_i') \cdot y_t}{\sum_{i=1}^{N_b} K(\cdot,x_i) + \sum_{i=1}^{N_p} K(\cdot,x_i')},
\end{split}
\end{equation}
where $\phi_t(\cdot) \in \mathbb{R} $ denotes the predictive probability output of $f'(\cdot;\theta)$ for the target class $t$. Meanwhile, $x_i' = (1-m_0) \odot x_i + T(m_0 \odot x)$ represents the poisoned samples in the training data and $y_i$ signifies the true label of the sample $x_i$. $K(x,x_i) = e^{-2\gamma||x-x_i||^2} (\gamma > 0)$. Since the training dataset of size $N_b$ is uniformly sampled from the $M$ different classes, it follows that there are an equal number of benign samples per class, with $\frac{N_b}{M}$ samples belonging to the $y_t$. Without loss of generality, we choose to keep the same settings as study \cite{r32}, assuming the target label $y_t = 1$ while others are 0. Then the regression solution can be transformed as follows:
\begin{equation}
\small
\begin{split}
\phi_t(\cdot) = \frac{\sum_{i=1}^{N_b / M} K(\cdot,x_i) + \sum_{i=1}^{N_p} K(\cdot,x_i')}{\sum_{i=1}^{N_b} K(\cdot,x_i) + \sum_{i=1}^{N_p} K(\cdot,x_i')}.
\end{split}
\end{equation}

When $N_p$ nears $N_b$, meaning the poisoning rate is nearly 50\%, the attacker can reach the optimal attack efficacy \cite{r6,r7,r9,r32}. Therefore we set $N_p$ = $N_b$.
We first test the predictive probability output using the poisoned sample $x_0' = (1-m_0) \odot x + T(m_0 \odot x)$ with the same trigger position in the training phase. The process is illustrated in Eq. (\ref{equation 14}).
\begin{equation}
\begin{split}
\small
&\sum_{i=1}^{N_p} K(x_0',x_i') - \sum_{i=1}^{N_b} K(x_0',x_i) = \sum_{i=1}^{N_p} K(x_0',x_i') - K(x_0',x_i)\\
&= \sum_{i=1}^{N_p} e^{-2\gamma||(1-m_0) \odot x + T(m_0 \odot x) - (1-m_0) \odot x_i - T(m_0 \odot x_i)||^2}\\
&- e^{-2\gamma||(1-m_0) \odot x + T(m_0 \odot x) - x_i||^2}\\
&= \sum_{i=1}^{N_p} e^{-2\gamma||(1-m_0) \odot (x-x_i)||^2} (e^{-2\gamma||m_0 \odot (T(x) - T(x_i))||^2}\\
&- e^{-2\gamma||m_0 \odot (T(x) - x_i)||^2})\\
&= \sum_{i=1}^{N_p} e^{-2\gamma||(1-m_0) \odot (x-x_i)||^2} (K(m_0 \odot T(x) , m_0 \odot T(x_i))\\
&- K(m_0 \odot T(x) , m_0 \odot x_i)).
\end{split}
\label{equation 14}
\end{equation}

The benign samples $x$ utilized to generate the poisoned samples $x_0'$ in the inference phase must not belong to the target class $y_t$. Otherwise the attacker has no incentive to introduce triggers in the first place. Therefore, $K(m_0 \odot T(x) , m_0 \odot T(x_i)) \gg K(m_0 \odot T(x) , m_0 \odot x_i)$. It indicates that $\sum_{i=1}^{N_p} K(x_0',x_i') > \sum_{i=1}^{N_b} K(x_0',x_i)$, which suggests $\phi_t(x_0') > 0.5$ and  $f'(x_0') = y_t$. This aligns with our expected outcomes, demonstrating the backdoor's capability to effectively classify the poisoned samples into the intended target class. 

Furthermore, for these benign samples $x$, it is understood that they would not be classified into the target class $y_t$ unless the trigger is introduced, and we can get Eq. (\ref{equation 15}).
\begin{equation}
\begin{split}
\small
&\sum_{i=1}^{N_p} K(x,x_i') - \sum_{i=1}^{N_b} K(x,x_i)  = \sum_{i=1}^{N_p} K(x,x_i') - K(x,x_i) < 0\\
&\Rightarrow \sum_{i=1}^{N_p} e^{-2\gamma||(1-m_0) \odot (x-x_i)  + m_0 \odot (x - T(x_i))||^2}\\
&- e^{-2\gamma||(1-m_0) \odot (x - x_i) + m_0 \odot (x - x_i)||^2} < 0\\
&\Rightarrow \sum_{i=1}^{N_p} e^{-2\gamma||(1-m_0) \odot (x-x_i)||^2} (e^{-2\gamma||m_0 \odot (x - T(x_i))||^2}\\
&- e^{-2\gamma||m_0 \odot (x - x_i)||^2}) < 0.
\end{split}
\label{equation 15}
\end{equation}
In addition we can get:
\begin{equation}
\begin{split}
\small
&\frac{\sum_{i=1}^{N_b / M} K(x,x_i) + \sum_{i=1}^{N_p} K(x,x_i')}{\sum_{i=1}^{N_b} K(x,x_i) + \sum_{i=1}^{N_p} K(x,x_i')} <0.5 \\
&\Rightarrow \sum_{i=1}^{N_b / M} K(x,x_i) < 0.5\sum_{i=1}^{N_b} K(x,x_i) - K(x,x_i')\\
&\Rightarrow \sum_{i=1}^{N_b / M} K(x,x_i) < 0.5\sum_{i=1}^{N_b} e^{-2\gamma||(1-m_0) \odot (x-x_i)||^2}\\
&\cdot(e^{-2\gamma||m_0 \odot (x - x_i)||^2}
- e^{-2\gamma||m_0 \odot (x - T(x_i))||^2}).
\end{split}
\label{equation 16}
\end{equation}

We then shift the trigger so that its new location no longer coincides with the original trigger region. We accomplish this process by adjusting the mask $m$, substituting the original mask $m_0$ with the updated mask $m_1$. For the new poisoned sample $x_1' = (1-m_1) \odot x + T(m_1 \odot x)$, we may assume that the backdoor is still in effect for it and can classify it into the target class $y_t$. So we can get Eq. (\ref{equation 17}).
It is obvious that the final form of Eq. (\ref{equation 16}) is highly similar to that of Eq. (\ref{equation 17}). Therefore, we further derive Eq. (\ref{equation 16}) and Eq. (\ref{equation 17}) to maintain formal consistency, leading to Eq. (\ref{equation 18}) and Eq. (\ref{equation 19}). 
\begin{figure*}[!t]
\centering
\begin{equation}
\small
\begin{split}
&\frac{\sum_{i=1}^{N_b / M} K(x_1',x_i) + \sum_{i=1}^{N_p} K(x_1',x_i')}{\sum_{i=1}^{N_b} K(x_1',x_i) + \sum_{i=1}^{N_p} K(x_1',x_i')} >0.5 \Rightarrow \sum_{i=1}^{N_b / M} K(x_1',x_i) > 0.5\sum_{i=1}^{N_b} K(x_1',x_i) - K(x_1',x_i')\\
&\Rightarrow \sum_{i=1}^{N_b / M} K(x_1',x_i) > 0.5\sum_{i=1}^{N_b} K((1-m_1) \odot x + T(m_1 \odot x),x_i) - K((1-m_1) \odot x + T(m_1 \odot x),(1-m_0) \odot x_i + T(m_0 \odot x_i))\\
&\Rightarrow \sum_{i=1}^{N_b / M} K(x_1',x_i) > 0.5\sum_{i=1}^{N_b} e^{-2\gamma(||(1-m_0-m_1) \odot (x-x_i)||^2 + ||m_0 \odot (x-x_i)||^2 + ||m_1 \odot (T(x)-x_i)||^2)}\\
&-e^{-2\gamma(||(1-m_0-m_1) \odot (x-x_i)||^2 + ||m_0 \odot (x-T(x_i))||^2 + ||m_1 \odot (T(x)-x_i)||^2)}\\
&\Rightarrow \sum_{i=1}^{N_b / M} K(x_1',x_i) > 0.5\sum_{i=1}^{N_b} e^{-2\gamma||(1-m_0-m_1) \odot (x-x_i)||^2}e^{-2\gamma||m_1 \odot (T(x)-x_i)||^2} \cdot(e^{-2\gamma||m_0 \odot (x - x_i)||^2} - e^{-2\gamma||m_0 \odot (x - T(x_i))||^2})\\
&\Rightarrow \sum_{i=1}^{N_b / M} e^{-2\gamma||(1-m_1) \odot (x-x_i) + m_1\odot (T(x) -x_i)||^2} > 0.5\sum_{i=1}^{N_b} e^{-2\gamma||(1-m_0-m_1) \odot (x-x_i)||^2}\\
&\cdot e^{-2\gamma||m_1 \odot (T(x)-x_i)||^2}(e^{-2\gamma||m_0 \odot (x - x_i)||^2} - e^{-2\gamma||m_0 \odot (x - T(x_i))||^2}).
\end{split}
\label{equation 17}
\end{equation}
\vspace{-8mm}
\end{figure*}
\begin{equation}
\begin{split}
\small
&\sum_{i=1}^{N_b / M} K(x,x_i) < 0.5\sum_{i=1}^{N_b} e^{-2\gamma||(1-m_0) \odot (x-x_i)||^2} \\
&\cdot(e^{-2\gamma||m_0 \odot (x - x_i)||^2} - e^{-2\gamma||m_0 \odot (x - T(x_i))||^2})\\
&\Rightarrow \sum_{i=1}^{N_b / M} e^{-2\gamma||(1-m_1) \odot (x-x_i)||^2}e^{-2\gamma||m_1\odot (x -x_i)||^2}\\
&< 0.5\sum_{i=1}^{N_b} e^{-2\gamma||(1-m_0-m_1) \odot (x-x_i)||^2} e^{-2\gamma||m_1 \odot (x-x_i)||^2}\\
&\cdot(e^{-2\gamma||m_0 \odot (x - x_i)||^2} - e^{-2\gamma||m_0 \odot (x - T(x_i))||^2}).
\end{split}
\label{equation 18}
\end{equation}
\begin{equation}
\begin{split}
\small
&\sum_{i=1}^{N_b / M} e^{-2\gamma||(1-m_1) \odot (x-x_i) + m_1\odot (T(x) -x_i)||^2}\\
&> 0.5\sum_{i=1}^{N_b} e^{-2\gamma||(1-m_0-m_1) \odot (x-x_i)||^2} e^{-2\gamma||m_1 \odot (T(x)-x_i)||^2}\\
&\cdot(e^{-2\gamma||m_0 \odot (x - x_i)||^2} - e^{-2\gamma||m_0 \odot (x - T(x_i))||^2})\\
&\Rightarrow \sum_{i=1}^{N_b / M} e^{-2\gamma||(1-m_1) \odot (x-x_i)||^2}e^{-2\gamma||m_1\odot (T(x) -x_i)||^2}\\
&> 0.5\sum_{i=1}^{N_b} e^{-2\gamma||(1-m_0-m_1) \odot (x-x_i)||^2} e^{-2\gamma||m_1 \odot (T(x)-x_i)||^2}\\
&\cdot(e^{-2\gamma||m_0 \odot (x - x_i)||^2} - e^{-2\gamma||m_0 \odot (x - T(x_i))||^2}).
\end{split}
\label{equation 19}
\end{equation}

We observe that Eq. (\ref{equation 18}) and Eq. (\ref{equation 19}) differ in only item $e^{-2\gamma||m_1\odot (x -x_i)||^2} , e^{-2\gamma||m_1\odot (T(x) -x_i)||^2}$ on both the left and right sides, yet their amplitude symbols are starkly contradictory.
As previously mentioned, the triggers we introduced are  invisible, resulting in the poisoned samples being nearly indistinguishable from the unaltered ones.
Thus, $e^{-2\gamma||m_1\odot (x -x_i)||^2} \approx e^{-2\gamma||m_1\odot (T(x) -x_i)||^2}$, indicating that the
magnitude signs in Eq. (\ref{equation 18}) and Eq. (\ref{equation 19}) should, theoretically, align. This demonstrates that our assumption that the backdoor would remain effective on the new poisoned sample $x_1'$ is wrong. Thus we can obtain :
\begin{equation}
\begin{split}
\small
&\frac{\sum_{i=1}^{N_b / M} K(x_1',x_i) + \sum_{i=1}^{N_p} K(x_1',x_i')}{\sum_{i=1}^{N_b} K(x_1',x_i) + \sum_{i=1}^{N_p} K(x_1',x_i')} < 0.5.
\end{split}
\label{equation 20}
\end{equation}
Therefore, when the spatial location of the trigger is changed, the poisoned sample will not be classified into the target class, at which point the backdoor ceases to be ineffective.

\end{proof}

{
\small
\bibliography{main}

\begin{thebibliography}{10}
\providecommand{\url}[1]{#1}
\csname url@samestyle\endcsname
\providecommand{\newblock}{\relax}
\providecommand{\bibinfo}[2]{#2}
\providecommand{\BIBentrySTDinterwordspacing}{\spaceskip=0pt\relax}
\providecommand{\BIBentryALTinterwordstretchfactor}{4}
\providecommand{\BIBentryALTinterwordspacing}{\spaceskip=\fontdimen2\font plus
\BIBentryALTinterwordstretchfactor\fontdimen3\font minus \fontdimen4\font\relax}
\providecommand{\BIBforeignlanguage}[2]{{%
\expandafter\ifx\csname l@#1\endcsname\relax
\typeout{** WARNING: IEEEtran.bst: No hyphenation pattern has been}%
\typeout{** loaded for the language `#1'. Using the pattern for}%
\typeout{** the default language instead.}%
\else
\language=\csname l@#1\endcsname
\fi
#2}}
\providecommand{\BIBdecl}{\relax}
\BIBdecl

\bibitem{r35}
J.~Zhang, M.~Z.~A. Bhuiyan, X.~Yang, T.~Wang, X.~Xu, T.~Hayajneh, and F.~Khan, ``Anticoncealer: Reliable detection of adversary concealed behaviors in edgeai-assisted iot,'' \emph{IEEE Internet of Things Journal}, 2021.

\bibitem{r36}
N.~Cai, D.~Wang, M.~Z.~A. Bhuiyan, L.~Han, and G.~Li, ``Lightsca: Lightweight side-channel attack via discrete cosine transform and residual networks,'' in \emph{Proceedings of the IEEE International Conference on High Performance Computing \& Communications; International Conference on Data Science \& Systems; International Conference on Smart City; International Conference on Dependability in Sensor, Cloud \& Big Data Systems \& Applications (HPCC/DSS/SmartCity/DependSys)}, 2022.

\bibitem{r1}
J.~Su, D.~V. Vargas, and K.~Sakurai, ``One pixel attack for fooling deep neural networks,'' \emph{IEEE Transactions on Evolutionary Computation}, 2019.

\bibitem{r2}
A.~Madry, A.~Makelov, L.~Schmidt \emph{et~al.}, ``Towards deep learning models resistant to adversarial attacks.'' in \emph{Proceedings of the International Conference on Learning Representations}, 2018.

\bibitem{r3}
W.~R. Huang, J.~Geiping, L.~Fowl \emph{et~al.}, ``Metapoison: Practical general-purpose clean-label data poisoning,'' \emph{Advances in Neural Information Processing Systems}, 2020.

\bibitem{r4}
A.~Shafahi, W.~R. Huang, M.~Najibi \emph{et~al.}, ``Poison frogs! targeted clean-label poisoning attacks on neural networks,'' \emph{Advances in Neural Information Processing Systems}, 2018.

\bibitem{r10}
H.~Chen, Y.~Gao, A.~Zhang \emph{et~al.}, ``Investigating the backdoor on dnns based on recolorization and reconstruction: From a multi-channel perspective,'' \emph{IEEE Transactions on Information Forensics and Security}, 2024.

\bibitem{r5}
T.~A. Nguyen and A.~T. Tran, ``Wanet - imperceptible warping-based backdoor attack,'' in \emph{Proceedings of the International Conference on Learning Representations}, 2021.

\bibitem{r6}
T.~Gu, K.~Liu, B.~Dolan-Gavitt \emph{et~al.}, ``Badnets: Evaluating backdooring attacks on deep neural networks,'' \emph{IEEE Access}, 2019.

\bibitem{r7}
Y.~Li, Y.~Li, B.~Wu \emph{et~al.}, ``Invisible backdoor attack with sample-specific triggers,'' in \emph{Proceedings of the IEEE/CVF International Conference on Computer Vision}, 2021.

\bibitem{r8}
J.~Bai, B.~Wu, Y.~Zhang \emph{et~al.}, ``\BIBforeignlanguage{en-US}{Targeted attack against deep neural networks via flipping limited weight bits},'' in \emph{\BIBforeignlanguage{en-US}{Proceedings of the International Conference on Learning Representations}}, 2021.

\bibitem{r9}
Y.~Liu, S.~Ma, Y.~Aafer \emph{et~al.}, ``Trojaning attack on neural networks,'' in \emph{Proceedings of the Annual Network And Distributed System Security Symposium}, 2018.

\bibitem{r11}
M.~Xue, C.~He, J.~Wang \emph{et~al.}, ``One-to-n \& n-to-one: Two advanced backdoor attacks against deep learning models,'' \emph{IEEE Transactions on Dependable and Secure Computing}, 2020.

\bibitem{r12}
M.~Xue, S.~Ni, Y.~Wu \emph{et~al.}, ``Imperceptible and multi-channel backdoor attack,'' \emph{Applied Intelligence}, 2024.

\bibitem{r13}
K.~D. Doan, Y.~Lao, and P.~Li, ``Marksman backdoor: Backdoor attacks with arbitrary target class,'' \emph{Advances in Neural Information Processing Systems}, 2022.

\bibitem{r14}
B.~Schneider, N.~Lukas, and F.~Kerschbaum, ``Universal backdoor attacks,'' in \emph{Proceedings of the International Conference on Learning Representations}, 2024.

\bibitem{r15}
X.~Chen, C.~Liu, B.~Li \emph{et~al.}, ``Targeted backdoor attacks on deep learning systems using data poisoning,'' \emph{arXiv preprint arXiv:1712.05526}, 2017.

\bibitem{r16}
Y.~Liu, X.~Ma, J.~Bailey \emph{et~al.}, ``Reflection backdoor: A natural backdoor attack on deep neural networks,'' in \emph{Proceedings of the European Conference on Computer Vision}, 2020.

\bibitem{r17}
Y.~Feng, B.~Ma, J.~Zhang \emph{et~al.}, ``Fiba: Frequency-injection based backdoor attack in medical image analysis,'' in \emph{Proceedings of the IEEE/CVF Conference on Computer Vision and Pattern Recognition}, 2022.

\bibitem{r18}
T.~Wang, Y.~Yao, F.~Xu \emph{et~al.}, ``An invisible black-box backdoor attack through frequency domain,'' in \emph{Proceedings of the European Conference on Computer Vision}, 2022.

\bibitem{r19}
K.~Doan, Y.~Lao, W.~Zhao \emph{et~al.}, ``Lira: Learnable, imperceptible and robust backdoor attacks,'' in \emph{Proceedings of the IEEE/CVF International Conference on Computer Vision}, 2021.

\bibitem{r24}
Z.~Zhang, Q.~Liu, Z.~Wang \emph{et~al.}, ``Backdoor defense via deconfounded representation learning,'' in \emph{Proceedings of the IEEE/CVF Conference on Computer Vision and Pattern Recognition}, 2023.

\bibitem{r28}
Y.~Liu, A.~Mondal, A.~Chakraborty \emph{et~al.}, ``A survey on neural trojans,'' in \emph{Proceedings of the International Symposium on Quality Electronic Design}, 2020.

\bibitem{r25}
B.~Chen, W.~Carvalho, N.~Baracaldo \emph{et~al.}, ``Detecting backdoor attacks on deep neural networks by activation clustering,'' \emph{arXiv preprint arXiv:1811.03728}, 2018.

\bibitem{r26}
H.~Chen, C.~Fu, J.~Zhao \emph{et~al.}, ``\BIBforeignlanguage{en-US}{Deepinspect: A black-box trojan detection and mitigation framework for deep neural networks},'' in \emph{\BIBforeignlanguage{en-US}{Proceedings of the International Joint Conference on Artificial Intelligence}}, 2019.

\bibitem{r34}
Y.~Zeng, W.~Park, Z.~M. Mao, and R.~Jia, ``Rethinking the backdoor attacks' triggers: A frequency perspective,'' in \emph{Proceedings of the IEEE/CVF International Conference on Computer Vision}, 2021.

\bibitem{r20}
R.~R. Selvaraju, M.~Cogswell, A.~Das \emph{et~al.}, ``Grad-cam: Visual explanations from deep networks via gradient-based localization,'' in \emph{Proceedings of the IEEE/CVF International Conference on Computer Vision}, 2017.

\bibitem{r21}
B.~Wang, Y.~Yao, S.~Shan \emph{et~al.}, ``Neural cleanse: Identifying and mitigating backdoor attacks in neural networks,'' in \emph{Proceedings of the IEEE Symposium on Security and Privacy}, 2019.

\bibitem{r22}
K.~Liu, B.~Dolan-Gavitt, and S.~Garg, ``Fine-pruning: Defending against backdooring attacks on deep neural networks,'' in \emph{Proceedings of the International Symposium on Research in Attacks, Intrusions, and Defenses}, 2018.

\bibitem{r23}
Y.~Gao, C.~Xu, D.~Wang \emph{et~al.}, ``Strip: A defence against trojan attacks on deep neural networks,'' in \emph{Proceedings of the Annual Computer Security Applications Conference}, 2019.

\bibitem{r33}
Y.~Gao, H.~Chen, P.~Sun \emph{et~al.}, ``Energy-based backdoor defense without task-specific samples and model retraining,'' in \emph{Proceedings of the International Conference on Machine Learning}, 2024.

\bibitem{r27}
Y.~Li, T.~Zhai, B.~Wu \emph{et~al.}, ``Rethinking the trigger of backdoor attack,'' \emph{arXiv preprint arXiv:2004.04692}, 2020.

\bibitem{r30}
A.~Jacot, F.~Gabriel, and C.~Hongler, ``Neural tangent kernel: Convergence and generalization in neural networks,'' \emph{Advances in Neural Information Processing Systems}, 2018.

\bibitem{r29}
Y.~Gao, H.~Chen, P.~Sun \emph{et~al.}, ``A dual stealthy backdoor: From both spatial and frequency perspectives,'' in \emph{Proceedings of the AAAI Conference on Artificial Intelligence}, 2024.

\bibitem{r31}
J.~Guo, A.~Li, and C.~Liu, ``{AEVA}: Black-box backdoor detection using adversarial extreme value analysis,'' in \emph{Proceedings of the International Conference on Learning Representations}, 2022.

\bibitem{r32}
J.~Guo, Y.~Li, X.~Chen \emph{et~al.}, ``{SCALE}-{UP}: An efficient black-box input-level backdoor detection via analyzing scaled prediction consistency,'' in \emph{Proceedings of the International Conference on Learning Representations}, 2023.

\end{thebibliography}
}

\vfill

\end{document}